\documentclass[english,twocolumn, prb, showpacs,superscriptaddress]{revtex4-1}
\usepackage[]{fontenc}
\usepackage[latin9]{inputenc}
\usepackage{color}
\usepackage{array}
\usepackage{amsmath}
\usepackage{amssymb}
\usepackage{graphicx}
\usepackage{dcolumn}
\usepackage{epstopdf}

\usepackage{color}

\newcommand{\red}{\color{red}}

\makeatletter

\providecommand{\tabularnewline}{\\}

\@ifundefined{textcolor}{}
{%
 \definecolor{BLACK}{gray}{0}
 \definecolor{WHITE}{gray}{1}
 \definecolor{RED}{rgb}{1,0,0}
 \definecolor{GREEN}{rgb}{0,1,0}
 \definecolor{BLUE}{rgb}{0,0,1}
 \definecolor{CYAN}{cmyk}{1,0,0,0}
 \definecolor{MAGENTA}{cmyk}{0,1,0,0}
 \definecolor{YELLOW}{cmyk}{0,0,1,0}
 }

\makeatother

\begin{document}

\title{ \textit{Ab initio} theory of electron-phonon mediated ultrafast spin relaxation
of laser-excited hot electrons in transition-metal
ferromagnets}

\author{K.\ Carva}
\email[]{karel.carva@fysik.uu.se}
\affiliation{Department of Physics and Astronomy, Uppsala University, Box 516,
S-75120 Uppsala, Sweden}
\affiliation{Charles University, Faculty of Mathematics and Physics, Department
of Condensed Matter Physics, Ke Karlovu 5, CZ-12116 Prague 2, Czech
Republic}
\author{M.\ Battiato}
\affiliation{Department of Physics and Astronomy, Uppsala University, Box 516,
S-75120 Uppsala, Sweden}

\author{D.\ Legut}
\affiliation{Nanotechnology Centre, VSB-Technical University of Ostrava, 17.\ listopadu 15, CZ-708 33~Ostrava, Czech Republic}
\author{P.\ M.\ Oppeneer}
\affiliation{Department of Physics and Astronomy, Uppsala University, Box 516,
S-75120 Uppsala, Sweden}
\date{\today}

\begin{abstract}
We report a computational theoretical investigation of electron spin-flip scattering induced by the electron-phonon interaction
in the %cubic %3$d$ 
 transition-metal ferromagnets bcc Fe, fcc Co and fcc Ni. The Elliott-Yafet electron-phonon spin-flip scattering is computed
from first-principles, employing a generalized spin-flip Eliashberg function as well as {\it ab initio} computed phonon dispersions.
Aiming at investigating the amount of electron-phonon mediated demagnetization in femtosecond laser-excited ferromagnets, the formalism is 
extended to treat laser-created thermalized as well as nonequilibrium, nonthermal hot electron distributions. 
Using the developed formalism we compute the phonon-induced spin lifetimes of hot electrons in Fe, Co, and Ni. The electron-phonon mediated 
demagnetization rate is evaluated for laser-created thermalized and nonequilibrium electron distributions. Nonthermal distributions are found to 
lead to a stronger demagnetization rate than hot, thermalized distributions, yet their demagnetizing effect is not enough to explain the
experimentally observed demagnetization occurring in the subpicosecond regime. 
% We compare our \textit{ab initio} calculated phonon-mediated %demagnetization rates 
% {\cyan spin-flip probabilities} with results obtained using the Elliott approximation. 
%{\cyan The Elliott spin-flip probability overestimates the {\it ab initio} electron-phonon spin-flip probability by a factor of two to four. KC: not so important and already discussed in PRL}
\end{abstract}

\pacs{78.47.J-, 78.20.Ls, 78.70.Dm, 78.20.Bh}

\maketitle

\section{Introduction}

In recent years it has been demonstrated that 
magnetization can be changed without applying an external magnetic field in extremely short timescales of the order of hundreds of femtoseconds.  \cite{r_96_Beaur_UltraNiDynam,r_05_Kimel_ultra_nontherm_exp, r_10_Kiril_UltraReview,r_09_Bigot_CohFemtoDemag}
Although it initial appeared that this would not be possible using strong pulsed magnetic fields,\cite{r_04_Tudos_UltSwitchLimit} it was
discovered that an ultrafast demagnetization of ferromagnetic transition metals could be induced by a femtosecond laser pulse. \cite{r_96_Beaur_UltraNiDynam, r_97_Hohlf_NiDyn_onlyTemp, r_98_Ju, r_00_Regen} 
 A closely related but
more complex process of optically induced magnetization switching has recently been discovered for ferrimagnetic systems with two antiparallel sublattice magnetizations. \cite{r_09_Vahap_Kalas_fsMORevers, r_11_Radu_TransMediat_AF, r_12_Ostle_UltraHeat_ReverFerri}
These discoveries offer possible routes
 to manipulate the magnetic moment on a sub-picosecond timescale and may lead to technological breakthrough in future ultrafast memory devices.

The observation of ultrafast all-optical demagnetization in elemental ferromagnets has led to an intensive and on-going debate on what actually is the
microscopic origin of the ultrafast process. \cite{Bovensiepen09, r_09_Mulle_SpinPol_HM_Femto, r_09_Zhang_TRMOKEpara, r_09_Bigot_CohFemtoDemag, r_10_Koopm_EYDemagInclGd, r_11_cbo_MOcontrov}
A first proposal for a microscopic explanation was based on direct transfer of angular momentum from the light, assisted by the
spin-orbit (SO) interaction. \cite{zhang00} Later it was however argued quantitatively that this source of  photon
angular momentum 
is insufficient to cause such a huge observed demagnetization, \cite{r_03_Koopm_SepMOArtifacts} when taking
into account the amount of photons present in the experiment and the estimated probability of a spin-flip (SF) excitation.
It has subsequently been argued that the ultrafast magneto-optical Kerr effect (MOKE) response on the femtosecond timescale could be modified by existing nonequilibrium (NEQ) electron distribution created by the fs pump laser,\cite{r_00_Koopm_StateFill, r_02_Kampf_UltraMO_Fe, r_03_Koopm_SepMOArtifacts, r_04_Oppen_NiDemag} yet this effect would disappear in a few hundred femtoseconds\cite{r_02_Guido_CoPt3_MOValidity} and the MOKE signal would thereupon follow the time-evolution of the magnetization dynamics. 
Hereafter a number of theoretical models have been proposed to explain the ultrafast demagnetization;  most of these are based on the assumption of a particular spin dissipation channel.\cite{r_05_Koopm_MicroModel_MagDynam, r_08_Carpe_ElMagnon_Fe, r_09_Krauss_CoNiComp_Coulomb, r_10_Koopm_EYDemagInclGd,  r_09_Zhang_TRMOKEpara, r_09_Bigot_CohFemtoDemag}
Ultrafast spin dissipation channels that have been put forward are Elliott-Yafet electron-phonon SF scattering, \cite{r_05_Koopm_MicroModel_MagDynam,r_10_Koopm_EYDemagInclGd} electron-magnon SF scattering,\cite{r_08_Carpe_ElMagnon_Fe,r_10_Schmi_Zhukov_UltraMagnon_FeOnCu} and electron-electron SF scattering; \cite{r_09_Krauss_CoNiComp_Coulomb}
 spin-orbit interaction is again the precursor in these electron--quasiparticle scatterings. 
Also direct, laser-induced SF processes\cite{r_09_Zhang_TRMOKEpara} and relativistic spin--light interaction\cite{r_09_Bigot_CohFemtoDemag} have been suggested.  Further, a distinct model which does not assume ultrafast spin-flips, but instead fast, superdiffusive transport of spins carried away with hot electrons has been proposed.\cite{r_10_bco_SpinTrDemag,r_12_bco_UltraSuperDiff_Layered} A few first observations\cite{r_11_Melni_UltraCarriersAuFe, Rudolf12, Vodungbo12, Pfau12} of laser-induced spin transport have been reported recently.

Even when the origin of the ultrafast spin relaxation is not known, model simulations on the basis of the 
Landau-Lifshitz-Gilbert and Landau-Lifshitz-Bloch equations can provide valuable insight. 
\cite{r_07_Atxit_MicroMD_LLB, r_08_Kazan_MultiModel_FePt, r_09_Vahap_Kalas_fsMORevers, r_10_Atxit_ThermMech_fsDemag, r_12_Ostle_UltraHeat_ReverFerri}
In these simulations suitably chosen longitudinal SF dissipation and transversal damping parameters, due to an unspecified microscopic mechanism, are adopted. With well-chosen dissipation parameters the measured laser-induced magnetization response can be captured, for elemental ferromagnets \cite{r_10_Atxit_ThermMech_fsDemag, Atxitia11} and for ferrimagnets with two sublattice magnetizations.\cite{r_11_Radu_TransMediat_AF, r_12_Ostle_UltraHeat_ReverFerri, Vahaplar12, r_12_Menti_Hellsv_SpinDyn_MultiSubL}
The achieved correspondence with the measured ultrafast spin-dynamics  may have implications for unveiling the ultrafast spin-flip channel.
For example, on the basis of Landau-Lifshitz-Bloch simulations it was recently argued, for the laser-induced demagnetization in Gd, that phonon-mediated spin-flips are needed above the Debye temperature, whereas below it electron-electron SF interaction should be sufficient.\cite{r_12_Sulta_Atxit_ElPhon_Gd}

%A viable modeling approach appears nonetheless to assume a spin-relaxation time due to an unspecified microscopic mechanism and perform equilibrium atomistic simulations of the laser-induced magnetization dynamics.\cite{Djordjevic07,atxitia07, atxitia10, atxitia11, Ostler12, Vahaplar12}
%{atxitia07, atxitia10, atxitia11, Ostler12, Vahaplar12}

Phonon-mediated Elliott-Yafet SF scattering has recently received significant attention.\cite{r_05_Koopm_MicroModel_MagDynam,r_10_Koopm_EYDemagInclGd}  On the basis of the Elliott-Yafet theory (in the Elliott spin-mixing approximation \cite{r_54_Elliot_SpinRelax_SO})  a unification of the theory of ultrafast demagnetization with the theory of Gilbert damping has been suggested.\cite{r_05_Koopm_DynModel_UnifGilbert,r_10_Fahnle_GilbertAndDemag,r_11_Faehn_UltraFastDissipMagD}
Moreover, strong arguments in favor of the Elliott-Yafet SF scattering as dominant mechanism for ultrafast laser-induced demagnetization 
were presented by Koopmans \textit{et al.} \cite{r_10_Koopm_EYDemagInclGd} Experimentally fitted SF probabilities were in reasonably good agreement with theoretical SF probabilities that were computed \textit{ab initio} using the Elliott spin-mixing approximation.\cite{r_09_Steia_Fahnl_EYcalcul} A large effective demagnetization was found when these SF probabilities were employed in the microscopic three temperature model (M3TM). \cite{r_10_Koopm_EYDemagInclGd} 
Elliott-Yafet  SF scattering combined with the M3TM has furthermore predicted correctly the observed relation between the demagnetization
rates in Ni, Co and Gd, \cite{r_10_Koopm_EYDemagInclGd} but these calculations rest on a number of approximations and contain a fitting parameter for each calculated metal. The measured temperature dependence of the laser-induced magnetization dynamics was consistent with the phonon-mediated spin-dissipation. \cite{r_12_Roth_TempDepDemagNi_Mechanism} 
{%\red \sout{A treatment of the ultrafast magnetic phase transition in FeRh on the basis of the Elliott-Yafet theory was proposed recently.\cite{r_11_Sandr_femtoFeRh_FP}}}

Although these results might provide support for the Elliott-Yafet phonon-mediated processes as mechanism underlying ultrafast demagnetization,
first-principles investigations have to be performed to quantify more precisely the amount of demagnetization that can be achieved by electron-phonon SF scattering. Recently, the first \textit{ab initio} investigations have been undertaken, \cite{r_11_cbo_EYSF, r_11_Essert_EPScat_Demag} yet more such investigations, considering different materials are required to obtain a complete picture.

To calculate the demagnetization rate created by  Elliott-Yafet electron-phonon processes, two major steps have to be performed. 
The first one is to evaluate the SF probability during a scattering event, which is assumed to be the origin
of the spin magnetization dissipation.
This step is clear from the theory point of view and the difference between methods of various groups
lies mainly in how detailed the description of scattering is, e.g., at the level of inclusion of real phonon dispersions and evaluation of the electron-phonon matrix elements (cf.\ Refs.\ \onlinecite{r_09_Steia_Fahnl_EYcalcul, r_11_cbo_EYSF, r_11_Essert_EPScat_Demag, r_11_Faehn_UltraFastDissipMagD}).
The second step is the calculation of the actual demagnetization
rate employing the calculated SF probabilities. In this step several assumptions may come into play.
One of these is the treatment of the laser-excited electron and spin systems and their thermalization. 
On the initial femtosecond timescale after irradiation the electron system is not in equilibrium
with neither the lattice nor the spin system. 
Note that the laser irradiation must be very intensive to cause observable demagnetization, hence a significant amount of electrons is excited during the process to high energy levels. The proper inclusion of this nonequilibrium situation might well be crucial for correctly modeling the ultrafast demagnetization. The occurring NEQ  distributions also represents
a complication for measurements of the spin dynamics on the fs timescale, which are commonly performed utilizing MOKE\cite{r_03_Koopm_SepMOArtifacts, r_09_Vorak_MEdge, r_12_Mathi_ExchIntTime_NiFe}
or the x-ray magnetic circular dichroism (XMCD),\cite{r_07_Stamm_FemtoNi,r_11_Radu_TransMediat_AF,r_11_Wiets_HotEl_SpinLat_Gd_Tb} as  the redistribution of electrons has to be taken into account 
 when interpreting experimental data. \cite{r_00_Koopm_StateFill,r_04_Oppen_NiDemag,r_09_clo_NiDemag,r_11_cbo_MOcontrov}
The laser-generated electron and spin distributions have, however, been modeled in rather different ways recently.\cite{r_09_Steia_Fahnl_EYcalcul, r_10_Koopm_EYDemagInclGd, r_11_cbo_EYSF, r_11_Essert_EPScat_Demag} Sometimes only a thermalized electron distribution has been assumed,\cite{r_10_Koopm_EYDemagInclGd} or the presence of laser-excited states has been approximated by averaging over a large energy interval.\cite{r_09_Steia_Fahnl_EYcalcul}    

%\sout{Also, sharply peaked structures of the spin-polarized density of states (DOS) that are typical for each of the transition metals have sometimes crudely been assumed to be constant.\cite{r_10_Koopm_EYDemagInclGd}}

Here we aim at developing a theoretical treatment for accurately calculating electron-phonon generated ultrafast spin relaxation {without employing the approximate Elliott relation}.
%Previously accurate calculations of the electron-phonon SF rate have performed on the basis of the SF Eliashberg function for Al in low-temperature equilibrium.\cite{r_98_Fabia_SpinRelCalc, r_99_Fabia_Sarma_PhIndSpinRel}
Our treatment 
%for laser-irradiated ferromagnets 
is based on a generalized spin- and energy- dependent Eliashberg function \cite{r_98_Fabia_SpinRelCalc}
and involves a new quantity, the SF probability as a function of electron energy. The computational scheme has been implemented in a relativistic \textit{ab initio} band-structure code, and hence it does not rely on assumptions regarding the shape of the spin-polarized density of states (DOS).
% or how it changes with time. Electron-phonon matrix elements are calculated using the frozen phonon approach.
Importantly, our computational approach is valid not only
for thermalized electron distributions, but also for the more general nonthermal distributions that are expected to exist in the material  within the first 300 fs after the pump pulse. Using the developed formalism we investigate the phonon-induced SF rates and spin dynamics of the ferromagnetic transition metals Fe, Co, and Ni, treating both thermalized as well as nonthermalized hot electron distributions.
Our calculated demagnetization rates shed promptly more light on the mechanism of ultrafast magnetization dynamics. 
In particular, we find that nonthermal electron distributions lead to a stronger demagnetization rate than thermal electron distributions, however, the contribution of phonon-induced SF scattering is not sufficient to explain the measured ultrafast demagnetization.

\section{Theory}

Within electron band theory, spin non-conserving processes in ferromagnetic crystalline solids arise from the spin-orbit 
interaction.  %Treat the Bloch states in the presence of spin-orbit interaction.
In the presence of the SO interaction the majority and minority Bloch  eigenstates $|\Psi_{\boldsymbol{k}n}^{\uparrow}\rangle$ and $|\Psi_{\boldsymbol{k}n}^{\downarrow}\rangle$
can be decomposed into spin-up and spin-down spinor parts,
%{\red Careful, what is the meaning of majority etc here}
\begin{equation}
|\Psi_{\boldsymbol{k}n}^{\uparrow}\rangle=a_{\boldsymbol{k}n}^{\uparrow}\left|\Uparrow\right\rangle +b_{\boldsymbol{k}n}^{\uparrow}\left|\Downarrow\right\rangle ,\; |\Psi_{\boldsymbol{k}n}^{\downarrow} \rangle=a_{\boldsymbol{k}n}^{\downarrow}\left|\Downarrow\right\rangle +b_{\boldsymbol{k}n}^{\downarrow}\left|\Uparrow\right\rangle .
\label{eq:EigDecomp}
\end{equation}
The 
%minority 
spinor components $b_{\boldsymbol{k}n}^{\sigma}$ ($\sigma$=$\uparrow,\downarrow$) are generally small (compared to $a_{\boldsymbol{k}n}^{\sigma}$)
and nonzero only if SO coupling is present. They represent
the degree of SO-induced spin-mixing.  

In the following we first describe the theory for electron-phonon generated spin-flip scattering in thermal equilibrium at low temperatures. 
Subsequently we extend the formalism to treat SF scattering for situations out of the low-temperature equilibrium.

\subsection{Phonon induced spin-flips in equilibrium}

An accurate calculation of the electron-phonon SF scattering has to be based on 
the phonon spectrum $\omega_{\boldsymbol{q} \nu}$ and the electron-phonon
matrix elements \cite{r_81_Grimvall_ElPhon} $g_{\boldsymbol{k}n,\boldsymbol{k}'n'}^{\nu} (\boldsymbol{q} )$.
Here $\nu$ and $\boldsymbol{q}$ denote the phonon mode and wavevector and $\boldsymbol{k}n$, $\boldsymbol{k}'n'$ are the electron quantum numbers;
momentum conservation demands $\boldsymbol{q}=\boldsymbol{k}' - \boldsymbol{k}$. The (squared)
spin-resolved electron-phonon matrix elements $g_{\boldsymbol{k}n,\boldsymbol{k}'n'}^{\nu\sigma \sigma'} (\boldsymbol{q} )$
are defined by
\begin{equation}
g_{\boldsymbol{k}n,\boldsymbol{k}'n'}^{\nu\sigma\sigma'} (\boldsymbol{q} )=|\boldsymbol{u}_{\boldsymbol{q} \nu}\cdot \langle\Psi_{\boldsymbol{k}n}^{\sigma} | \nabla_{\boldsymbol{R}}V| \Psi_{\boldsymbol{k}' n'}^{\sigma'} \rangle |^{2}\,,\label{eq:ElPhonElem}
\end{equation}
where $V$ is the total potential felt by the electrons, $\boldsymbol{u}_{\boldsymbol{q} \nu}$
is the phonon polarization vector, and $\nabla_{\boldsymbol{R}}$ denotes
the gradient with respect to the displacements of the atoms,\cite{r_81_Grimvall_ElPhon,r_86_Lam_SC_ElPhon}
which correspond to the mode $\boldsymbol{u}_{\boldsymbol{q} \nu}$. 

%{\red
%Several contributions to electron-phonon SF scattering can be distinguished.  As noted by Elliott \cite{r_54_Elliot_SpinRelax_SO} even the spin-diagonal part of  $V$ can connect eigenstates of majority and minority spin because of the SO spin-mixing
%present in the eigenstates, thus effectively allowing for spin-flip scattering. A further contribution to SF processes stems from the small spin-nondiagonal part of $V$, due to SO coupling.
 %\cite{r_63_Yafet_SpinLatRel,r_53_Overh_MagRelax} 
 %}

To derive a suitable low temperature formulation we consider which electronic states $| \Psi_{\boldsymbol{k} n}^{\sigma} \rangle$
can participate in the scattering. Energy conservation in the {electron-phonon scattering}
dictates the condition $\delta(E_{\boldsymbol{k}'{n}'}^{\sigma'} -E_{\boldsymbol{k}n}^{\sigma}-\hbar\omega_{\boldsymbol{q}\nu})$. 
At zero temperature electronic states up to the Fermi energy $E_{F}$
are occupied, therefore  conduction electrons with energies $E_{\boldsymbol{k}n}^{\sigma}$
in the range $[E_{F}-\hbar\Omega,\ E_{F}]$ are allowed to absorb
a phonon with energy $\Omega$. The energy of the final state $E_{\boldsymbol{k}'n'}^{\sigma'}$,
then lies in the range $[E_{F},\ E_{F}+\hbar\Omega]$. If one can
neglect the difference in the electronic states that differ by an energy $\Delta$
that is smaller than the maximum phonon energy $\hbar\omega_{max}$ 
(i.e., $E_{\boldsymbol{k}n}^{\sigma}$ and $E_{\boldsymbol{k}n}^{\sigma}+\Delta$) a suitable approximation can be made.
It is customary to describe
the distribution of states with energy $E$ participating in the scattering
by a $\delta$-function broadened by a width $\Delta$, denoted by $\tilde{\delta}(E-E_{F})$.
Such broadening is also needed for numerical implementation of formulas
involving $\delta$-functions.
% in this manner. 
Note that maximal phonon frequencies 
are typical of about 35 meV and the broadening of the electron distribution is already 25 meV at room temperature. 
Consequently, we 
can write the condition for the initial and final state together with the
requirement for energy conservation in a symmetrical form: $\tilde{\delta}(E_{{\boldsymbol{k}}n}^{\sigma}-E_{F})\tilde{\delta}(E_{{\boldsymbol{k}'}n'}^{\sigma'}-E_{F})$.
The approximation to ignore the distinction between conduction electron states differing by less than the maximum phonon energy was already made
in earlier works.\cite{r_81_Grimvall_ElPhon,r_99_Fabia_Sarma_PhIndSpinRel} 

The next step to describe the electron-phonon scattering is to introduce the equilibrium Eliashberg function.\cite{r_81_Grimvall_ElPhon} 
The standard, spin-diagonal equilibrium Eliashberg function is obtained when the squared electron-phonon matrix elements 
are integrated over all possible initial and final electronic states, restricted by the phonon energy $(\Omega=\omega_{{\boldsymbol{q}}\nu})$
as a parameter. \cite{r_81_Grimvall_ElPhon}
% the equilibrium Eliashberg function \cite{r_81_Grimvall_ElPhon} is obtained. 
For our purpose it is necessary to define a spin-dependent equilibrium
Eliashberg function $\alpha_{\sigma \sigma'}^{2}F^{0}(\Omega )$,
resolved with respect to the spin state of initial and final states, which
reads 
\begin{widetext} 
\begin{equation}
\!\!\!\!\alpha_{\sigma \sigma'}^{2}F^{0} (\Omega )=\frac{1}{2M\Omega}\sum_{\nu,n,n'}\int d\boldsymbol{k}\int d\boldsymbol{k}'g_{\boldsymbol{k}n,\boldsymbol{k}'n'}^{\nu\sigma \sigma'} (\boldsymbol{q}=\boldsymbol{k}' - \boldsymbol{k})\delta (\omega_{\boldsymbol{q}\nu}-\Omega )\tilde{\delta} (E_{\boldsymbol{k}n}^{\sigma}-E_{F} )\tilde{\delta}(E_{{\boldsymbol{k}'}n'}^{\sigma'}-E_{F}) .
\label{eq:Elbg0}
\end{equation}
\end{widetext} 
The spin-flip processes are given by the terms with
$\sigma$$\neq$$\sigma'$, hence, the equilibrium SF Eliashberg function is given by $\alpha_{\uparrow\downarrow}^{2}F^{0}(\Omega)$.
\cite{r_99_Fabia_Sarma_PhIndSpinRel} The sum over diagonal elements  $\sigma$=$\sigma'$
corresponds to the standard Eliashberg function, $\alpha^{2}F^{0}(\Omega)$. 
%{\red The approximation to ignore the distinction between states differing by less than the maximum phonon energy was already 
%in the early work of Eliashberg (ref) (see also Ref.\ \onlinecite{r_81_Grimvall_ElPhon}).}
%{\blue here or later: Note the equivalence of $\alpha_{\uparrow\downarrow}^{2}F^{0}(\Omega)=\alpha_{\downarrow\uparrow}^{2}F^{0}(\Omega)$,
%which is proved in the following subsection for a more general case.}
%{\red we should also mention/address the units ... P}
{Characteristic features of the electron-phonon scattering, which are important for the evaluation of Eq.\ (\ref{eq:Elbg0}), are the facts
that the change in electron energy is small, typically below 40 meV, but any momentum change in the scattering process is possible.}

As long as one is not interested in knowledge about electron-phonon scattering contributions
stemming from specific phonons it is useful to integrate over all phonon
energies and obtain the spin-resolved transition rate $w_{\sigma\sigma'}^{0}$,
\begin{equation}
w_{\sigma\sigma'}^{0}=\int_0^{\omega_{max}}  d\Omega\, \alpha_{\sigma \sigma'}^{2}F^{0} (\Omega )\left[1+2N (\Omega )\right]\,,
\label{trans-rate}
\end{equation}
where $N (\Omega )$
is the phononic Bose-Einstein distribution function.  Analogously, one can define the total transition rate $w^{0}$ 
(which in the current formulation includes \textit{both} spin-diagonal and nondiagonal scattering events).
Lastly, one can introduce the total SF probability
$P_{S}$ which is defined as the ratio of the SF and total scattering
rates,
\begin{equation}
P_{S}=\frac{w_{{S}}^{0}}{w^{0}}\,.
\label{eq:Equi-prob}
\end{equation}

%\subsection{Temporal evolution of spin population}
\subsection{Laser-irradiated ferromagnets}

To approach further the physical situation which is in the focus of the present study, 
we need to examine electron-phonon spin-flip scattering in laser-heated ferromagnets. 
Due to the pump-laser excitation and the subsequent process of electron thermalization, 
not only electron states in the immediate vicinity of the Fermi energy will be involved, 
also energetically deep-lying states that are reached by the pump laser as well as states at 
higher energies above $E_F$ that become populated have to be taken into account.

Following the above derivation, it is straightforward to define a generalized spin- and energy- dependent
Eliashberg function,
\begin{widetext}
\begin{equation}
\!\!\!\!\alpha_{\sigma \sigma'}^{2} {F} (E,\Omega )=\frac{1}{2M\Omega}\sum_{\nu,n,n'}\int d\boldsymbol{k}\int d\boldsymbol{k}' g_{\boldsymbol{k}n,\boldsymbol{k}'n'}^{\nu \sigma \sigma'} (\boldsymbol{q} ) \delta (\omega_{\boldsymbol{q} \nu}
- | \Omega | ) \delta (E_{\boldsymbol{k}n}^{\sigma}-E) \delta (E_{\boldsymbol{k}'n'}^{\sigma}-E_{\boldsymbol{k}n}^{\sigma'}-\hbar\Omega) \,.
\label{eq:gen-Eliashberg}
\end{equation}
\end{widetext} 
A negative $\Omega$ is possible and allowed (for absorption processes). As we are interested in
the regime of the order of hundreds of femtoseconds after the laser pulse, we can assume that 
the lattice has not yet been heated up and therefore the room temperature phonon distribution $N(\Omega )$
is assumed. This assumption is substantiated by recent measurements showing a rise of the lattice temperature 
of Ni in a few picoseconds after laser excitation, which is much slower than the ultrafast demagnetization. \cite{r_10_Wang_TempDepElPhon}

%As has been pointed out in several works {\red (citations!) KC: rather rewrite it:  
In a laser-pumped system the electrons are redistributed depending on the laser frequency and the amount of absorbed radiation. 
{The electronic system can be described by suitably modified distributions functions.\cite{r_00_Hohlf,r_04_Oppen_NiDemag, r_09_clo_NiDemag}}
{Also here} we describe this redistribution by band-index independent
occupation factors $f_{\sigma} (E)$, thus catching the
key quantities of the laser-pumped electron system -- its spin and energy dependence.
The assumption that all states labeled by $\sigma$ and $E$ have the same
occupancy is partially justified by the simple band structure of the
studied metals in the region above $E_{F}$. We note that it is possible to go
beyond this approximation with the presented formalism, but at the cost
of significant numerical complications.  Using this approximation
the spin-resolved transition rate\cite{r_63_Yafet_SpinLatRel}
that is the key quantity for magnetization
evolution is defined as 
\begin{eqnarray}
S^{\sigma\sigma'}\!\!=\!\!\iint d\Omega dE \, \alpha_{\sigma\sigma'}^{2} {F} (E,\Omega)f_{\sigma}(E) \times ~~~~~  \nonumber \\
%\, \,\, \,\,\,
(1\!-\!f_{\sigma'}(E\!+\!\hbar \Omega)) [\Theta(\Omega)\!+\!N(\Omega) ] .
\label{eq:SpinEvGen}
\end{eqnarray} 

The main goal of this study is to determine the total temporal evolution
of spin moment. In order to understand the relation between the spin evolution and 
typical electron distributions occurring after the laser pulse we reformulate the
above expressions using phonon-integrated quantities. This is done at
the cost of neglecting the difference between $f_{\sigma} (E )$
and $f_{\sigma} (E+\hbar\Omega)$, and the difference between
the electron-phonon matrix elements of states with energies  $E_{\boldsymbol{k}n}^{\sigma}$
and $E_{\boldsymbol{k}n}^{\sigma}+\hbar\Omega$. The latter approximation
is a completely analogy to the one made {above} in deriving the equilibrium Eliashberg
function $\alpha_{\sigma \sigma'}^{2}F^{0}(\Omega)$. It is a plausible
approximation because the phonon energy is much lower than the range of
electron energies made available due to the pump laser, or the energy
on which the band structure would be significantly changed. The results obtained
with this approximation have been checked against calculations using the accurate
Eq.\ (\ref{eq:SpinEvGen}) for the case of Ni. 

This approximation allows to neglect $\Omega$ in $\delta (E_{\boldsymbol{k}'n'}-E_{\boldsymbol{k}n}-\hbar\Omega)$,
when simultaneously the $\delta$-function is replaced by its broadened counterpart, $\tilde{\delta}$. This leads
to a reformulation of the Eliashberg function similar to the one made in the equilibrium case, specifically,
\begin{widetext}
\begin{equation}
\alpha_{\uparrow \downarrow}^{2} F (E,\Omega ) \simeq \frac{1}{2M |\Omega |}\sum_{\nu,n,n'}\int d\boldsymbol{k}\int d\boldsymbol{k}' g_{\boldsymbol{k}n,\boldsymbol{k}'n'}^{\nu\uparrow \downarrow} 
(\boldsymbol{q} )\delta (\omega_{\boldsymbol{q} \nu}- |\Omega | )\tilde{\delta} (E_{\boldsymbol{k}n}^{\uparrow}-E ) \tilde{\delta} (E_{\boldsymbol{k}'n'}^{\downarrow}-E ).
\label{eq:EnEliashApp}
\end{equation}

Next, to achieve a further re-writing of the equations, we investigate how the generalized Eliashberg functions for spin-majority to spin-minority scattering and \textit{vice versa} are related.
 To this end, we note that upon interchanging $\boldsymbol{k}\leftrightarrow \boldsymbol{k}'$ in the
integration and employing $g_{\boldsymbol{k}n,\boldsymbol{k}'n'}^{\nu\,\sigma \sigma'} (\boldsymbol{q} )=g_{\boldsymbol{k}'n',\boldsymbol{k}n}^{\nu\,\sigma' \sigma} (\boldsymbol{q} )$
we obtain
\begin{equation}
\alpha_{\uparrow \downarrow}^{2}F (E,\Omega )=\frac{1}{2M |\Omega |}\sum_{\nu,n,n'}\int d\boldsymbol{k}\int d\boldsymbol{k}' g_{\boldsymbol{k}n,\boldsymbol{k}'n'}^{\nu \downarrow \uparrow} (\boldsymbol{q} )\delta
(\omega_{\boldsymbol{q}\nu}- |\Omega | )\tilde{\delta} (E_{\boldsymbol{k}n}^{\downarrow}-E)\tilde{\delta} (E_{\boldsymbol{k}'n'}^{\uparrow}-E )\,=\alpha_{\downarrow \uparrow}^{2}F (E,\Omega ) .
\end{equation}
\end{widetext}
Hence, the equivalence $\alpha_{\uparrow \downarrow}^{2}F (E, \Omega) = \alpha_{\downarrow \uparrow}^{2}F (E, \Omega)$
is proven. The approximation to neglect the influence of $\Omega$ on the energies while broadening the $\delta$-function also elevates the need for distinguishing 
a negative $\Omega$, since the difference between phonon absorption
and emission becomes negligible and $\alpha_{\uparrow\downarrow}^{2}F (E,\Omega)\approx \alpha_{\uparrow\downarrow}^{2}F (E,-\Omega )$.

We can now define $w_{\sigma\sigma'} (E )$, which is a generalization of
the above-defined $w_{\sigma\sigma'}^{0}$, i.e., a spin- and energy-dependent
scattering rate that involves the average over all available states at
a given energy $E$ and all phonon states (or equivalently, all destination
states),
\begin{equation}
w_{\sigma\sigma'} (E) =\int_{0}^{\infty}d\Omega\, \alpha_{\sigma \sigma'}^{2}F (E,\Omega)\left[1 +2 N (\Omega)\right] .
\label{eq:SDRate_En}
\end{equation}
Consequently, 
the spin-resolved transition rates are 
\begin{equation}
S^{\sigma\sigma'}=\int dE\, w_{\sigma\sigma'} (E )f_{\sigma} (E )(1-f_{\sigma'} (E )) .
\label{eq:trans_rates}
\end{equation}
Apparently, we obtain $w_{\uparrow\downarrow} (E )=w_{\downarrow\uparrow} (E) = w_{S}(E)$,
because the same expression is valid for $\alpha_{\uparrow \downarrow}^{2}F (E,\Omega)$.
Therefore we can introduce the spin \textit{decreasing} transition rate $S^{-}=S^{\uparrow\downarrow}$
and the spin \textit{increasing} rate $S^{+}=S^{\downarrow\uparrow}$, which are given
by formulas that differ only through the occupation factors, 
\begin{eqnarray}
S^{-} \!\! & = &\!\! \int dE\, w_{S}(E)f_{\uparrow}(E)\left(1-f_{\downarrow} (E)\right), \nonumber \\
S^{+} \!\! & = & \!\! \int dE\, w_{S} (E)f_{\downarrow}(E)\left(1-f_{\uparrow} (E)\right) .
\label{eq:rates+-}
\end{eqnarray}

%\begin{equation}
%S^{\sigma\sigma'}\!\!\!=\!\!\iint\!\!\alpha_{\sigma\sigma'}^{2}F\!\left(E,\Omega\right)f_{\sigma}\!\left(E\right)\left(1\!-\! f_{\sigma'}\!\left(E\!+\!\Omega\right)\right)\left(\Theta\!\left(\Omega\right)\!+\! N\!\left(\Omega\right)\right)\! d\Omega dE
%\label{eq:SpinEvGen}
%\end{equation}
%where a negative $\Omega$ is allowed (for absorption), and the spin-
%and energy-dependent generalized Eliashberg function is defined as
%\begin{widetext}
%\begin{equation}
%\!\!\!\!\alpha_{\sigma \sigma'}^{2}F\left(E,\Omega\right)=\frac{1}{2M\Omega}\sum_{\nu,n,n'}\int d\mathbf{k}\int d\mathbf{k}'g_{\mathbf{k}n,\mathbf{k}'n'}^{\nu\sigma,\sigma'}\left(\mathbf{q}\right)\delta (\omega_{\mathbf{q\nu}}-\left|\Omega\right| )\delta (E_{\mathbf{k}n}^{\sigma}-E) \delta (E_{\mathbf{k'}n'}^{\sigma}-E_{\mathbf{k}n}^{\sigma'}-\hbar\Omega)\,.
%\label{gen-Eliashberg}
%\end{equation}
%\end{widetext} 
%Room temperature phonon distribution $N\!\left(\Omega\right)$
%is assumed in calculations, since we study the regime in the order
%of hundreds femtosecond after the laser pulse, when lattice temperature
%is not altered significantly. 

Employing this formulation the temporal evolution of the spin moment can be connected with the (nonequilibrium) 
electron distributions that are typically expected to occur in laser-excited ferromagnets.
The total electron-phonon scattering rate is given simply as $w (E)=\sum_{\sigma\sigma'}w_{\sigma\sigma'} (E)$.
The SF probability for an electron at a given energy $E$ during an 
%Elliott-Yafet 
electron-phonon scattering can be defined as  $p_{S}(E)=w_{S} (E)/w (E)$.
The total SF probability for a system with electron occupancies described by a distribution
$f_{\sigma} (E)$ is 
\begin{equation}
P_{S}=\left(S^{-}+S^{+}\right)/\sum_{\sigma\sigma'}S^{\sigma\sigma'} .
\label{eq:total-P}
\end{equation}
 Importantly, the crucial quantity for demagnetization is the demagnetization rate, 
which arises as the balance of spin-increasing and spin-decreasing spin-flip scatterings. It is
given by ${d}M/{d}t=2\mu_{B}\left(S^{-}-S^{+}\right)$, where $M$ is the $z$-component of the spin moment.
%{\red see appendix \ref{sec:SpinMomChange} for a detailed discussion. !?}
%
%To compare to experimentally determined demagnetization rates $\tau_M$ we employ the
%magnetization's time derivative $dM/dt$ at $t=0$. To do so, we assume the commonly used 
%experimental function describing superfast demagnetization, 
%$M(t) - M(0) = -M(0)(1 - e^{-t/{\tau_M}})$$\times e^{-t/{\tau_R}}$ (which also contains the re-magnetization time $\tau_R$), giving $dM/dt (0) = -M(0)/ \tau_M$.
Note that we consider here the initial demagnetization  $dM/dt (t=0)$ in a first order approximation, which allows us to compare to measured demagnetization rates, 
$dM/dt (0) = -[M(0)-M_{\rm min.}]/ \tau_M$, with $M_{\rm min.}$ the achieved minimal magnetization. A higher-order magnetization change induced by a reduction of the exchange field is not taken into account. 
 
 Lastly, we also define a relative quantity called demagnetization ratio, 
%independent of the total electron-phonon transition rate, but 
by taking the difference between
spin decreasing and increasing transition processes, but normalized to the total transition rate: 
\begin{equation}
D_{S}=\left(S^{-}-S^{+}\right)/\sum_{\sigma\sigma'}S^{\sigma\sigma'} .
\label{eq:D_S}
\end{equation}

\subsection{The Elliott relation}

As the evaluation of the spin-dependent electron-phonon matrix elements is a demanding computational step, several approximations
have been introduced in the past. 
One of these is the Elliott approximation, which we describe here in more detail as it was used in recent
theoretical works to explain the ultrafast laser-induced demagnetization\cite{r_09_Steia_Fahnl_EYcalcul,r_10_Koopm_EYDemagInclGd} 
 as well as the ultrafast magnetic phase transition occurring in FeRh.\cite{r_11_Sandr_femtoFeRh_FP} This approximation is not used in our model, only,  we have computed  some results based on it (denoted as Elliott SF probability), which are shown in graphs for sake of comparison.

Elliott \cite{r_54_Elliot_SpinRelax_SO}  originally pointed out that even the spin-diagonal part of the potential $V$ can connect
eigenstates of  majority and minority spin because of the spin-mixing
present in the eigenstates, thus effectively allowing for a spin-flip scattering. 
Making several approximations, Elliott could derive
a relation between the spin lifetime $\tau_{_{SF}}$ for a general kind of
scattering event which has a spin-diagonal lifetime $\tau$.
% can be derived \cite{r_98_Fabia_SpinRelCalc,
The employed  assumptions are that the material is a paramagnetic metal, that the  variations
of the electron-phonon matrix elements over the Brillouin zone (BZ) are small,  that $b_{\boldsymbol{k}n}$ is constant over the BZ, and $b_{\boldsymbol{k}n}^{\sigma}\ll a_{\boldsymbol{k}n}^{\sigma}$.
\cite{r_54_Elliot_SpinRelax_SO,r_98_Fabia_SpinRelCalc}
The resulting relation called after Elliott
employs the Fermi-surface  averaged spin-mixing of eigenstates,
\begin{equation}
\langle b^{2}\rangle =\sum_{\sigma,n}\int d\boldsymbol{k} \left|b_{\boldsymbol{k}n}^{\sigma}\right|^{2}\tilde{\delta} (E_{\boldsymbol{k}n}^{\sigma}-E_{F}),
\end{equation}
and predicts the SF probability $P_{S}^{b^{2}}$ to be approximately
\begin{equation}
P_{S}^{b^{2}}=\frac{\tau}{\tau_{_{SF}}}=4\left\langle b^{2}\right\rangle.
 \label{eq:EllRel}
\end{equation}

The Elliott relation can also be generalized to nonequilibrium situations.\cite{r_11_cbo_EYSF}
To this end we define a SF density of states as an averaged $b^{2}$-component of all states at a given energy $E$,
in analogy to the usual definition of the total DOS $n (E)$:
\begin{equation}
n_{\uparrow \downarrow} (E)=\sum_{\sigma, n}\int d\boldsymbol{k}\left|b_{\boldsymbol{k}n}^{\sigma}\right|^{2}\delta (E_{\boldsymbol{k}n}^{\sigma}-E) .
\label{eq:nb2_En}
\end{equation}
Applying the Elliott approximation for one electron at energy $E$ we obtain
an energy-resolved SF probability $p_{S}^{b^{2}} (E)=4 \langle b^{2} \rangle (E)=4n_{\uparrow \downarrow} (E )/ n (E)$.

Analogous to the expressions in the preceding subsection, the total Elliott SF probability $P_{S}^{b^2}$ of an electron system
after laser-excitation can be computed by adopting a representative electron distribution $f_{\sigma} (E)$ and performing an
energy integration similar to the one in  Eq.\ (\ref{eq:trans_rates}).   To this end,
$w_{\uparrow \downarrow} (E)$ has to be 
 replaced by $n_{\uparrow \downarrow} (E)$ and $w (E)$ by $n (E)$. 
 The total Elliott SF probability follows from an expression equivalent to Eq.\ (\ref{eq:total-P}).
 
 The Elliott relation was originally intended to treat electron-phonon scattering, but, because of the approximations made it does no longer rely on any phonon characteristics of the scattering. Therefore the Elliott SF probability
 would be the same irrespective if the SF scattering  is due to phonons or e.g., defects.
 
Apart from the spin-mixing in the wavefunctions a different SF scattering can arise from 
the spin-orbit coupling part $V_{SO}$ of the potential, as originally proposed by Overhauser.\cite{r_53_Overh_MagRelax}
 Yafet \cite{r_63_Yafet_SpinLatRel} showed that, at low temperatures, the contribution of this term to the 
 SF scattering almost cancels the spin-mixing contribution of Elliott (see also Ref.\ \onlinecite{r_99_Fabia_Sarma_PhIndSpinRel}).
%Experimental investigations\cite{r_78_Beune_ElliotRel_Metals}  for nonmagnetic metals confirmed Yafet's conjecture
%(see also Ref.\ \onlinecite{r_99_Fabia_Sarma_PhIndSpinRel}).
Notwithstanding, experimental investigations\cite{r_78_Beune_ElliotRel_Metals}  for nonmagnetic metals indicated 
that the trend given by Elliott's relation remained approximately valid, up to a multiplication with a materials specific constant within a variation of roughly one order of magnitude; but larger deviations were also reported for some metals.\cite{r_78_Beune_ElliotRel_Metals} 
%{\green what to do with this?  if it reduces the effect, it is not good that we don't have it -P}
 
An essential assumption made in the derivation of Elliott's relation, which is relevant for the discussion of ultrafast demagnetization in ferromagnets,
 is that the material treated is a nonmagnetic metal,
i.e., the scattering electron from initial state $| \boldsymbol{k}n \rangle$ can undergo a spin-flip and goes back to the 
same spin-degenerate $| \boldsymbol{k}n \rangle$ as final state.  In spite of this, the Elliott relation has  been applied to ferromagnetic
metals,\cite{r_09_Steia_Fahnl_EYcalcul,r_10_Koopm_EYDemagInclGd} but a recent investigation showed that it does fail for ferromagnetic metals with 
strongly exchange split bands. \cite{r_11_cbo_EYSF}
The obvious reason for this it that in exchange-split ferromagnetic bands the SF scattering electron has to go to
a different final band state which has a spin different from the original one.

 %
%The Elliott relation does not take into account the character of scattering
%involved, the spin-flip probability would be the same for scattering
%on defects or photons. 
%{\green What to do with the part below?\\}
%{\blue
%Note that the phononic character of scatterers
%immediately provides few important characteristics: the change of
%electron energy during scattering is very small, below $40$ meV,
%while any momentum change is possible. Furthermore the of scattering
%associated to a given phonon energy and momentum is selected by phononic
%distribution function - Bose-Einstein distribution $N\left(\Omega\right)$}

%Experimentally this relation was found to be valid up to a multiplication
%by a material specific constant with variation smaller than one order
%of magnitude \cite{r_78_Beune_ElliotRel_Metals} for various paramagnetic
%metals. Estimate for Ni in equilibrium was given using this formula:
%$p_{S}^{b^{2}}\sim0.1$ \cite{r_09_Steia_Fahnl_EYcalcul}.

%A further contribution to the SF scattering processes originates from the small 
%spin-nondiagonal part of $V$, due to the spin-orbit coupling. $\nabla_{R}V_{SO}$ is now often called
%the Yafet term,\cite{r_63_Yafet_SpinLatRel} although originaly proposed
%by Overhauser {\red Ref!}, and was recently evaluated for Ni.\cite{r_11_Essert_EPScat_Demag}. 

\subsection{Numerical implementation}

%{\red the part below has been worked through and the previous corrections are implemented -P}

The above derived equations for the electron-phonon SF scattering have been implemented
within a first-principles band structure code. 
The electronic structure calculations are based on the density functional theory (DFT) and 
performed with the local spin density approximation (LSDA).\cite{r_65_ks, Perdew92} 
The full-potential  linearized
augmented plane wave (FP-LAPW) method for the electronic structure calculations is used because of its ability
to describe phonons with high accuracy. 
We find that the calculated phonon dispersions of Ni and Fe are in
reasonable agreement with other \textit{ab initio} calculations and with experimental
data.\cite{r_00_DalC_Ni_Phonon_DFPT} 
The electron-phonon coupling elements
are calculated self-consistently\cite{r_86_Lam_SC_ElPhon} within
the ELK FP-LAPW code (http://elk.sourceforge.net/), hence $\nabla_{\boldsymbol{R}}V$
is evaluated for each $\boldsymbol{q}$ inside a supercell commensurate
with $\boldsymbol{q}$. The calculations of the spin-resolved equilibrium Eliashberg
function [Eq.\ (\ref{eq:Elbg0})] and its nonequilibrium counterpart  [Eq.\ (\ref{eq:EnEliashApp})], 
the energy-dependent SF rate as well as the Elliott spin-mixing ratio have been implemented 
 and are thus made available for the future study of a wide variety of metals. 
For systems with a 
complex band structure like transition metals and heavier elements
a good mesh density in reciprocal space is needed in order to obtain
good BZ averages. On the other hand, high $\boldsymbol{k}$-space density
of coupling elements requires big supercells, which makes calculations
numerically demanding even for simple Ni. Here we have used a 4$\times$4$\times$4 mesh of
phonon $\boldsymbol{q}$ points. As a test of our numerical implementation we have 
calculated the SF and non-SF Eliashberg function of Al in equilibrium.
These were found to be in good agreement with previous calculations,\cite{r_99_Fabia_Sarma_PhIndSpinRel}
with the SF Eliashberg function being approximately $10^5$ times smaller than the non-SF Eliashberg function.

\section{Results}

Our \textit{ab initio} computed results for the elemental $3d$ ferromagnets are presented in the following.
First, the  generalized spin- and energy- dependent Eliashberg functions and scattering rates are given.
For Ni we in addition compare our calculated results with those obtained by two different approximations:
the rigid-ion approximation by Nordheim \cite{r_31_Nordh_RION_Approx} and an approximation introduced by Wang \textit{et al.}\cite{r_94_Wang_EliashbApprox} 
Subsequently, we present computed electron-phonon induced spin lifetimes of hot, nonthermal electrons
and  our results for phonon-mediated demagnetization in laser-excited $3d$ ferromagnets.

\subsection{\textit{Ab initio} SF probabilities and SF scattering rates}

\subsubsection{fcc Ni}

\begin{figure}[t,b]
\includegraphics[width=0.80\columnwidth]{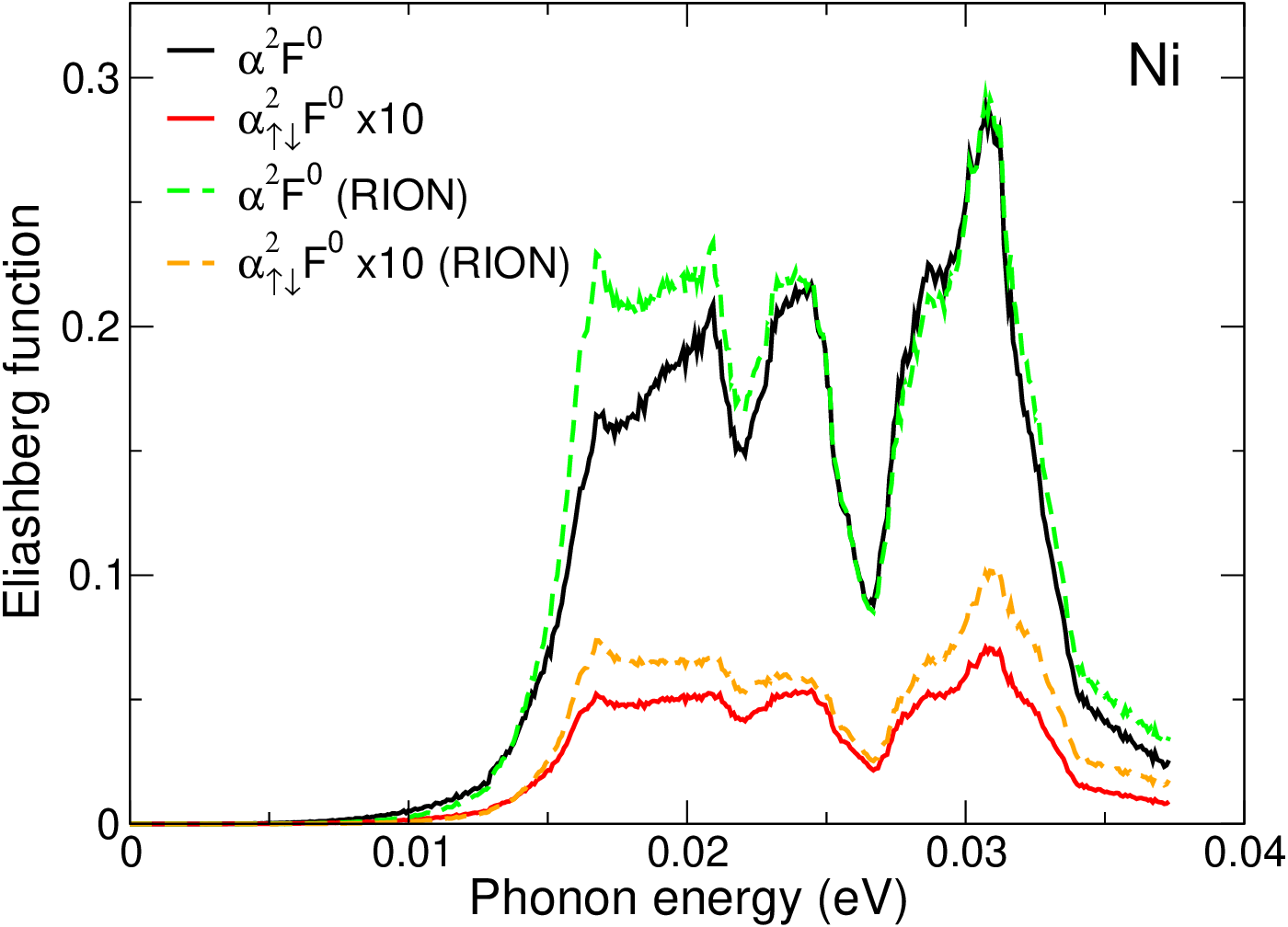}
\caption{
\label{fig:Elbg_SFElbg_Ni-1} 
(color online) \textit{Ab initio} calculated spin-flip Eliashberg function $\alpha_{\uparrow \downarrow}^2 F^0 (\Omega )$ 
and total Eliashberg function $\alpha^2 F^0 (\Omega )$  as function of phonon energy $\Omega$
for Ni in equilibrium. For comparison the SF and total Eliashberg functions computed
with the rigid ion approximation (labeled with RION) are also given.
}
\end{figure}

The key information for examining the electron-phonon spin-flip probability for Ni in equilibrium 
at low temperatures ($< 300$ K) 
is given by the equilibrium SF Eliashberg function (as compared to the total or non-SF Eliashberg function).
In Fig.\  \ref{fig:Elbg_SFElbg_Ni-1} we show the calculated SF and  total Eliashberg functions.
The SF Eliashberg function is about 50 times smaller than its non-SF counterpart.
As a result, {as will be discussed below,} the corresponding total SF probability $P_S$ in equilibrium Ni [Eq.\ (\ref{eq:Equi-prob})] is
of the order of $10^{-2}$.

In Fig.\ \ref{fig:Elbg_SFElbg_Ni-1} we include results computed with the rigid ion approximation.\cite{r_31_Nordh_RION_Approx}
In this approximation, the total electron-lattice potential $V (\boldsymbol{r}, \{ \boldsymbol{R}_i \} )$ is written as a superposition of on-site potentials $v (\boldsymbol{r} - \boldsymbol{R}_i )$ and for small core displacements it is assumed that the potential follows rigidly the motion of nuclei.\cite{r_31_Nordh_RION_Approx,r_81_Grimvall_ElPhon}
The rigid ion approximation is convenient to be used in conjunction with the atomic sphere approximation in band-structure calculations; recently, it was applied to study electron-phonon interaction is optically excited metals.\cite{r_11_Essert_EPScat_Demag} Here we observe that}
the correspondence between the unapproximated  Eliashberg functions and those obtained with 
the rigid-ion approximation is surprisingly good. All detailed structures in the functions are present, with at the most
a difference in their amplitude. This suggests that the rigid-ion approximation could be used to
achieve accurate SF probabilities. 

\begin{figure}[t,b]
	\includegraphics[clip,width=0.85\columnwidth]{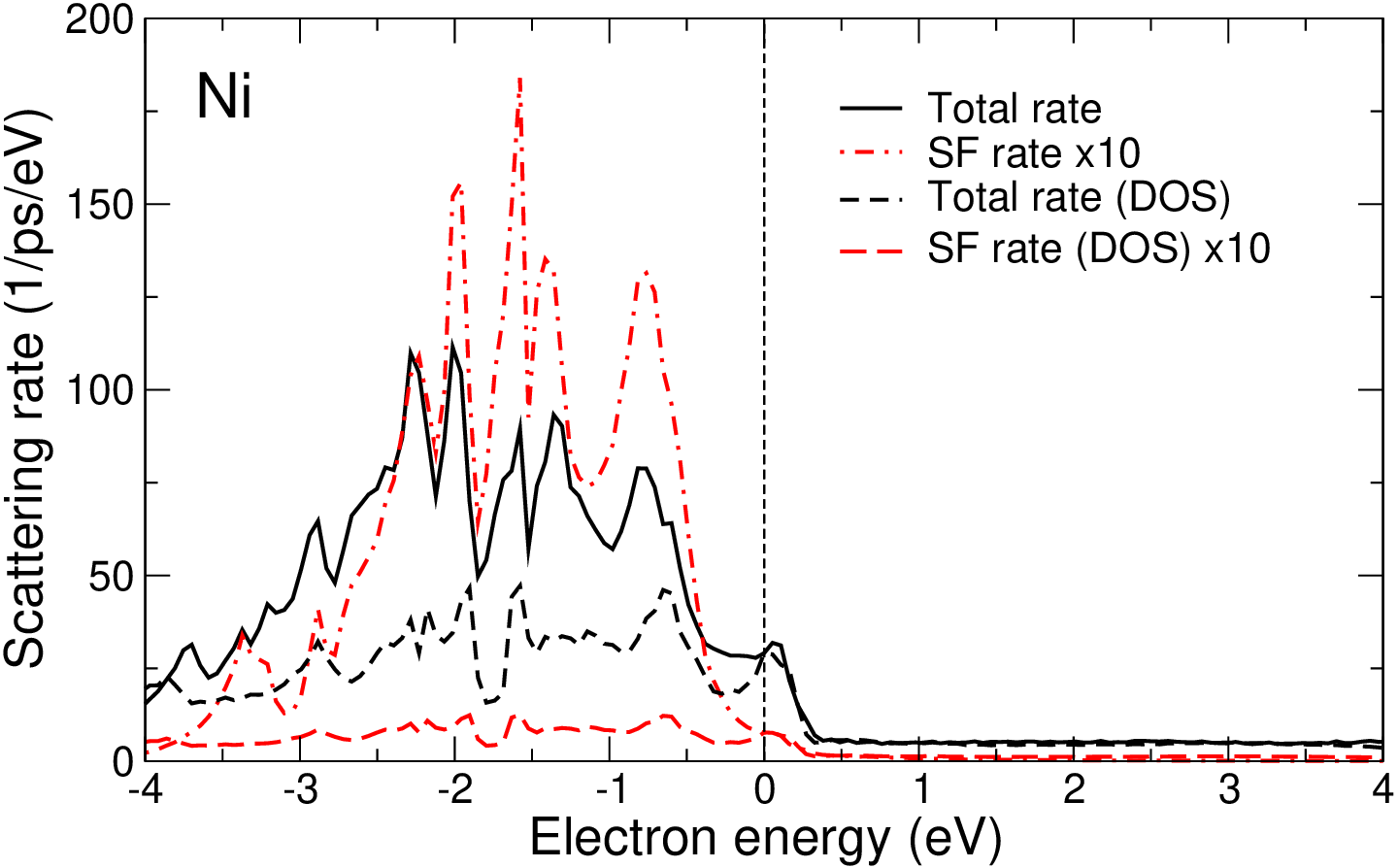}
\caption{\label{fig:En_SF} (color online)
 Energy-resolved electron-phonon total and
 SF scattering rates $w(E)$ and $w_{S}(E)$ for Ni obtained from direct \textit{ab initio} calculations.
For comparison their counterparts obtained from the Ni DOS employing the approximation of
Wang {\it et al.} \cite{r_94_Wang_EliashbApprox} are also shown. }
\end{figure}

To treat laser-heated ferromagnets we need to include in the study the behavior of electrons away from the 
Fermi level. We thus consider the energy-resolved SF and non-SF scattering rates, $w_{{S}} (E)$ and $w (E)$;
the calculated scattering rates are shown in Fig.\ \ref{fig:En_SF}.
Near the Fermi energy the phonon-induced SF rate is about 50 times smaller than the total rate, consistent 
with the behavior of the equilibrium SF and non-SF Eliashberg functions. At energies deeper below $E_F$ the SF scattering
rate is only up to 10 times smaller. Consequently, it can already be expected that, when electrons up to about  1.55 eV below $E_F$  
are affected by the laser-excitation, the effective SF probability can increase considerably beyond the equilibrium SF probability.

Wang {\it et al}. \cite{r_94_Wang_EliashbApprox} have proposed an
approximation to avoid the tedious calculation of the energy-dependent
Eliashberg function. 
%(or scattering rate 
This approximation entails using the equilibrium Eliashberg function
scaled by the  DOS, i.e.,
\begin{equation}
\alpha_{\sigma \sigma'}^{2}F (E,\Omega ) \approx \alpha_{\sigma \sigma'}^{2}F^0 (\Omega) \frac{n (E_{F} )}{n (E)}\,.
\label{eq:WangEliashberg}
\end{equation}
This approximation was used in a previous computational investigation 
of the  NEQ electron-phonon coupling  in Ni.\cite{r_08_Lin_ElPhon_Noneq}
% KC: Ni and other 7 metals
From our calculations of $\alpha_{\sigma \sigma'}^{2}F (E,\Omega )$ we can easily undertake
a verification of this approximation. 
The approximate energy-dependent Eliashberg function as predicted by Wang \textit{et al.}'s formula 
%(\ref{eq:WangEliashberg}) 
is shown, too, in Fig.\ \ref{fig:En_SF}. Around the Fermi energy the approximation is reasonable, 
as expected. However, at energies of more than 0.5 eV below $E_F$ the approximation becomes worse;
the total electron-phonon scattering rate is predicted to be about a factor 2 too small by Wang \textit{et al.}'s formula.
The use of this formula appears to be especially insufficient for studying the spin-flip scattering rate,
which is off by more than an order of magnitude.

The energy-dependent SF probability $p_S (E)$ as well as Elliott's SF probability
$p_S^{b^2} (E)$ have been computed \cite{r_11_cbo_EYSF} recently for Ni, and are therefore not shown here.
It was noted that both SF probabilities showed strong variations with electron energy, particular in the range of the $d$ bands.
Furthermore, the SF probability obtained from Elliott's relation could deviate a factor 
2 to 3 from the directly compute SF probability.
This deviation has been explained by the fact that Elliott's relation was originally
derived for paramagnetic metals, where the same states are equally available to both spins, and a spin-flipping electron 
can, in each $\boldsymbol{k}$-point, scatter back to the same state. 
This assumption is however only poorly fulfilled in ferromagnets with strongly exchange-split bands. \cite{r_11_cbo_EYSF}

\subsubsection{bcc Fe}

\begin{figure}
\includegraphics[width=0.80\columnwidth]{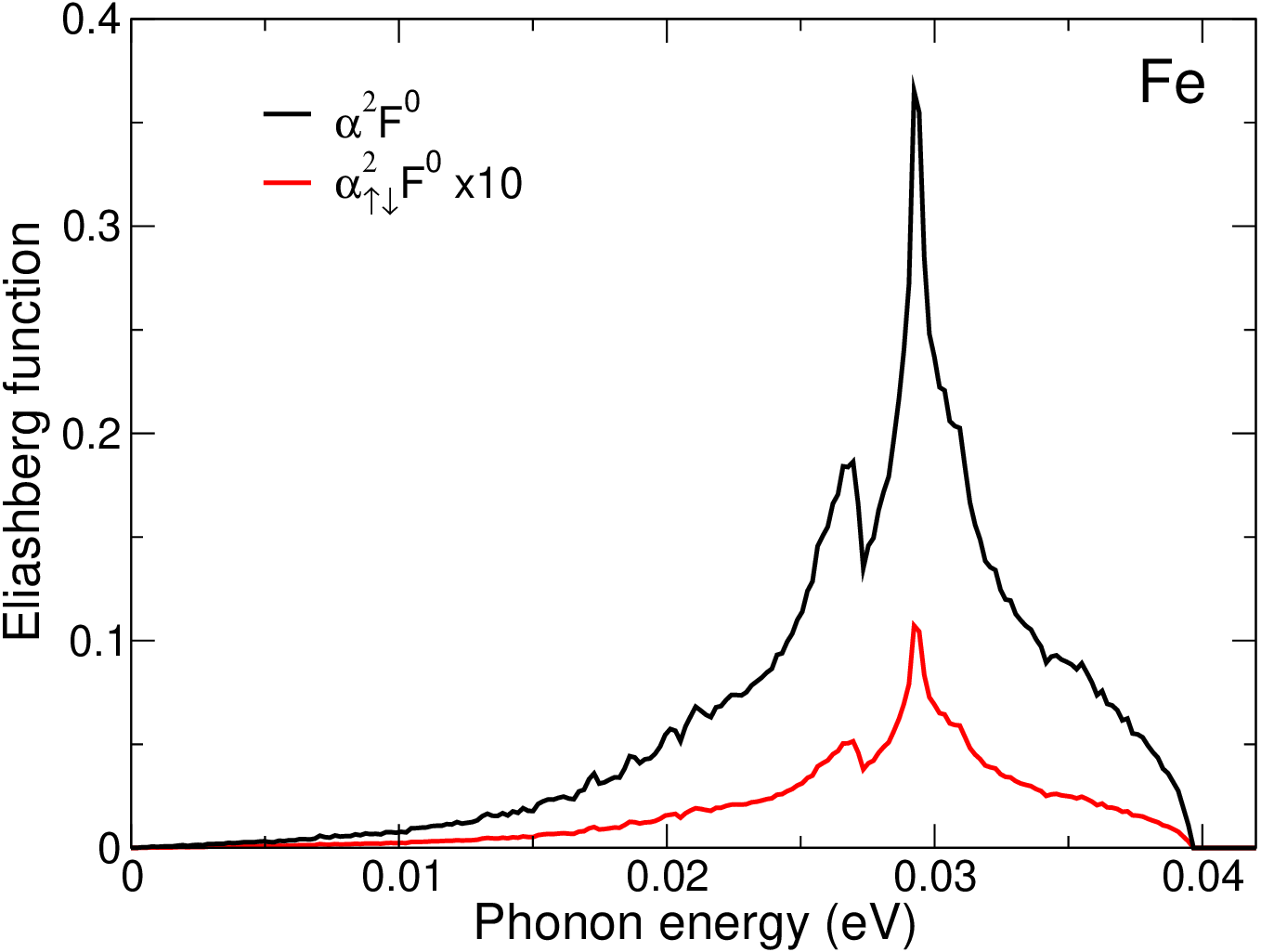}
\caption{(color online) 
\textit{Ab initio} calculated SF Eliashberg function $\alpha_{\uparrow \downarrow}^2 F^0 (\Omega )$ 
and total Eliashberg function $\alpha^2 F^0 (\Omega )$  of bcc Fe in equilibrium.
\label{fig:Eliash-Fe}
}
\end{figure}

The \textit{ab initio} calculated SF and total Eliashberg functions of bcc Fe
are shown in Fig.\ \ref{fig:Eliash-Fe}. 
Here, both the SF and total Eliashberg functions are more concentrated  in a narrow
peak area as compared to the case of Ni (cf.\ Fig.\ \ref{fig:Elbg_SFElbg_Ni-1}). 
The SF function is about 40 times smaller than the total Eliashberg function.

%{\red Where? Nonetheless, the ratio of integrals
%over these functions lead to a SF probability very similar to the
%one of Ni (Tab. \ref{tab:SFRates-1}).}

\begin{figure}
\includegraphics[width=0.85\columnwidth]{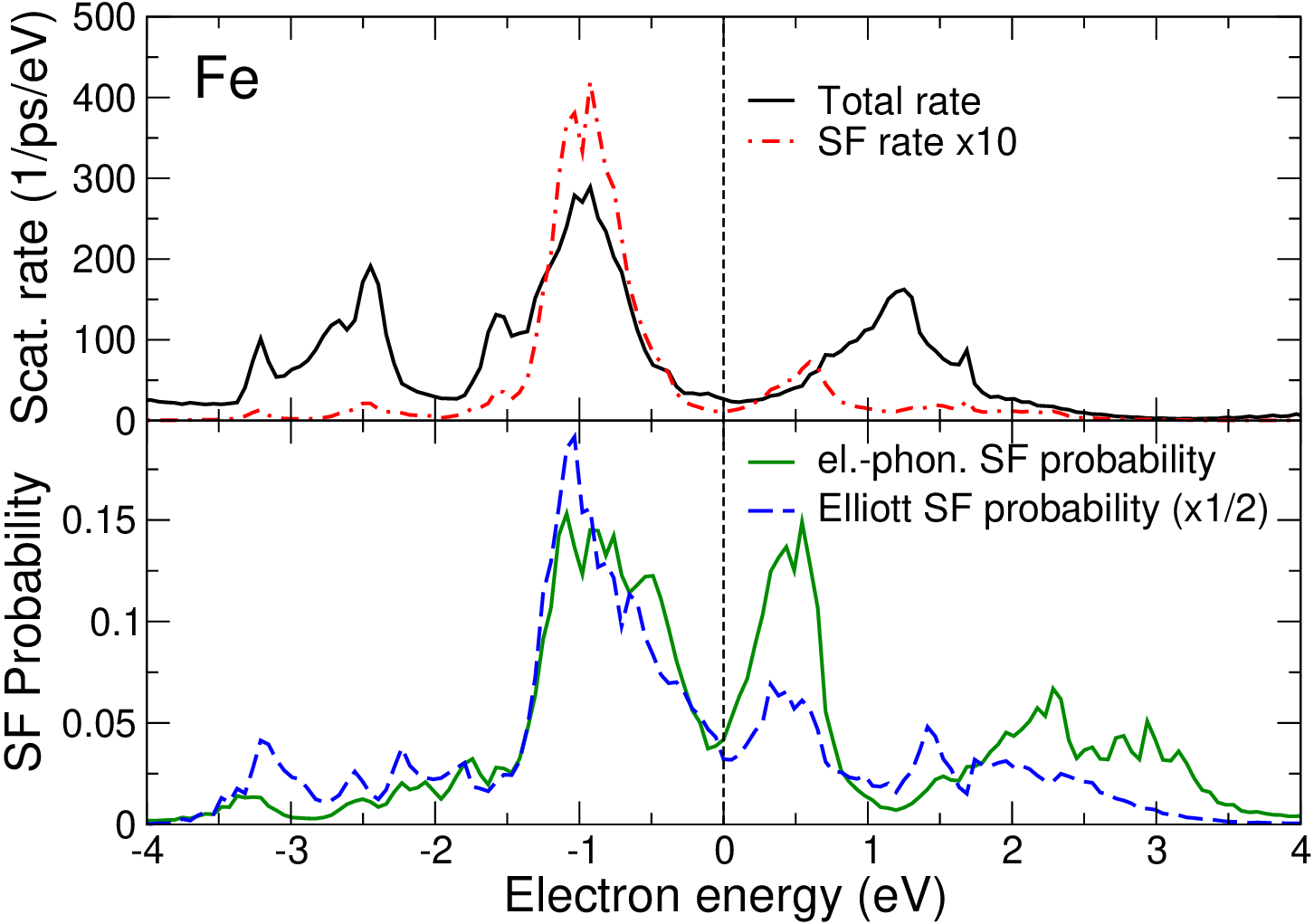}
\caption{(color online) 
Top: calculated energy-resolved  total and
SF electron-phonon scattering rates, $w(E)$ and $w_{S} (E)$, of Fe. 
Bottom panel: calculated energy-resolved SF probability
$p_{S} (E)$ and the approximate  SF probability 
$p_{S}^{b^{2}} (E)$ (divided by two) obtained from Elliott's relation.}
\label{fig:En-rates-Fe}
\end{figure}

The calculated energy-resolved electron-phonon SF and total scattering rates
% $w_S (E)$ and $w (E)$, 
are shown in Fig.\ \ref{fig:En-rates-Fe} (top). The SF scattering rates are again
about a factor of ten smaller than the non-SF scattering rates. In comparison to Ni both scattering rates
display more structure above $E_F$.
% Note the strong energy variations of w.
In Fig.\  \ref{fig:En-rates-Fe} (bottom) we show the computed electron-phonon
SF probability $p_S (E)$ as well as the SF probability $p_S^{b^2} (E)$  obtained from the
Elliott relation.
%Energy dependent scattering rates $w_{S}^{ep}\left(E\right)$ and
%SF probabilities $p_{S}^{ep}\left(E\right)$ shown in Fig. \ref{fig:En_SF}
These contain a few distinct features which differentiate Fe from Ni, namely
the presence of a SF probability peak above the Fermi level, with a value
of over 0.1. Also, the electron-phonon coupling is overall stronger in Fe than in
Ni in the relevant range of $\pm1.55$ eV around the Fermi level by a factor
of 2, while the difference of the SF probabilities of Fe and Ni at their Fermi levels  is small.
As all energy-dependent quantities vary strongly in the $-1.55$ eV to +1.55 eV region,
the use of only Fermi level quantities to compute laser-induced demagnetization
would be a very poor approximation.
It is also worthwhile to note that the SF probability $p_{S}^{b^{2}} (E)$
obtained with the Elliott approximation
is a factor of two larger than the directly computed SF probability for energies to $-2$  eV 
below the Fermi energy, whereas it is about the same size in the peak at 0.5 eV.

%\begin{figure}
%\includegraphics[width=0.85\columnwidth]{fig_Fe/Fe_SFRates}
%\caption{\label{fig:En-rates-Fe} (color online) 
%Energy-resolved electron-phonon total and
%spin-flip (SF) scattering rates $w^{ep}$ and $w_{S}^{ep}$, SF probability
%$p_{S}^{ep}$ and (approximate SF probability from Elliott relation
%$p_{SF}^{b^{2}}$ ) for Fe.}
%\end{figure}

\subsubsection{fcc Co}

\begin{figure}
\includegraphics[width=0.80\columnwidth]{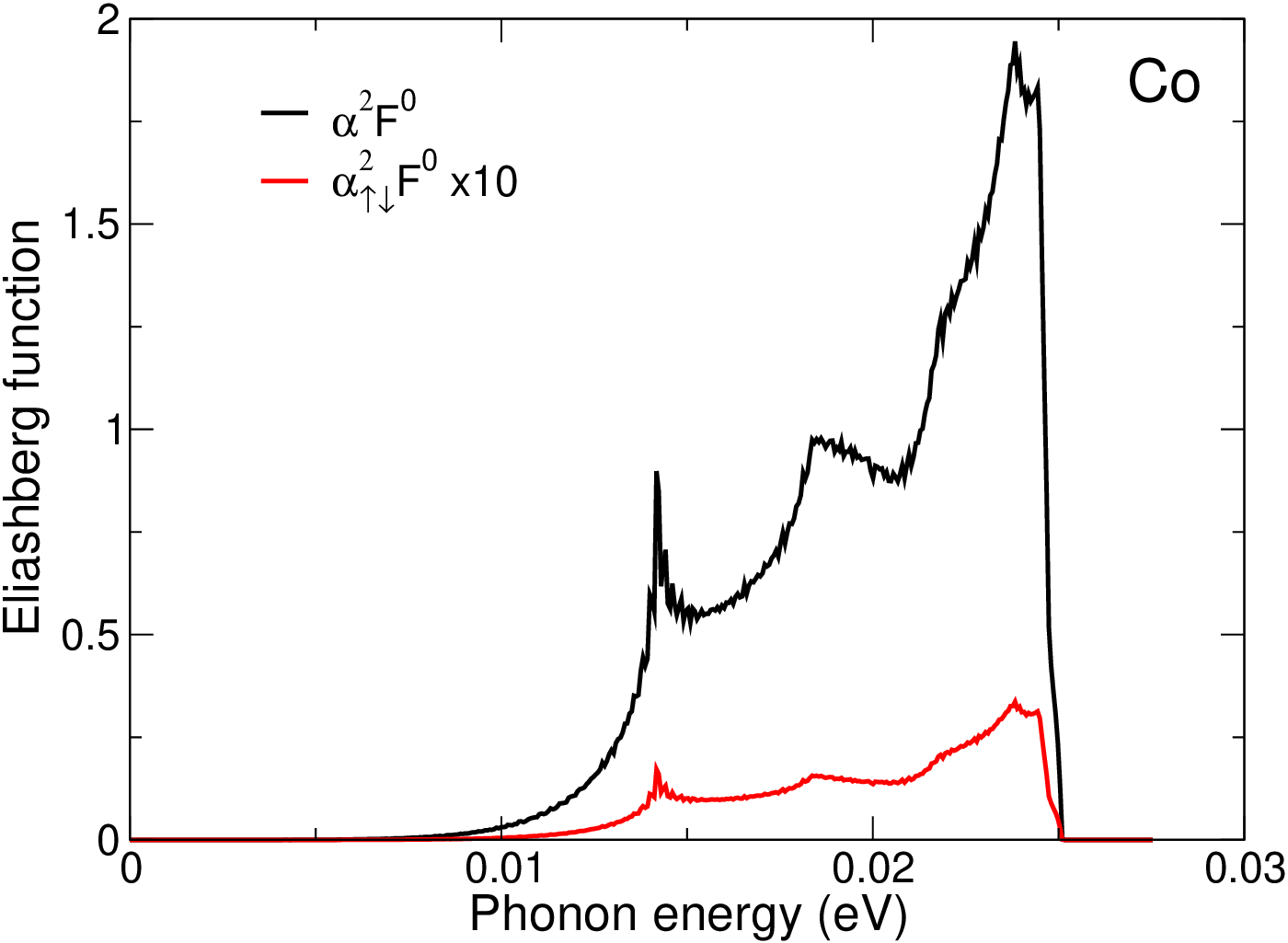}
\caption{\label{fig:Elbg-Co}
(color online) \textit{Ab initio} calculated SF and total Eliashberg functions of fcc Co in equilibrium,
as a function of the phonon energy. %{\red why is Co so much larger? -P }
}
\end{figure}

The \textit{ab initio} calculated SF and total Eliashberg functions of fcc Co
are shown in Fig.\ \ref{fig:Elbg-Co}. 
Both the total and SF Eliashberg functions of Co exhibit
overall higher values than those of the other two $3d$ ferromagnets, indicating a stronger
electron-phonon coupling just at the Fermi level. 
However, the ratio of the
 SF to the total Eliashberg function is somewhat lower.
 %, indicating a low SF probability
%$p_{s}$ (see Tab. \ref{tab:SFRates-1}), in agreement with the previous
%calculation \cite{r_09_Steia_Fahnl_EYcalcul}.

\begin{figure}
\includegraphics[width=0.85\columnwidth]{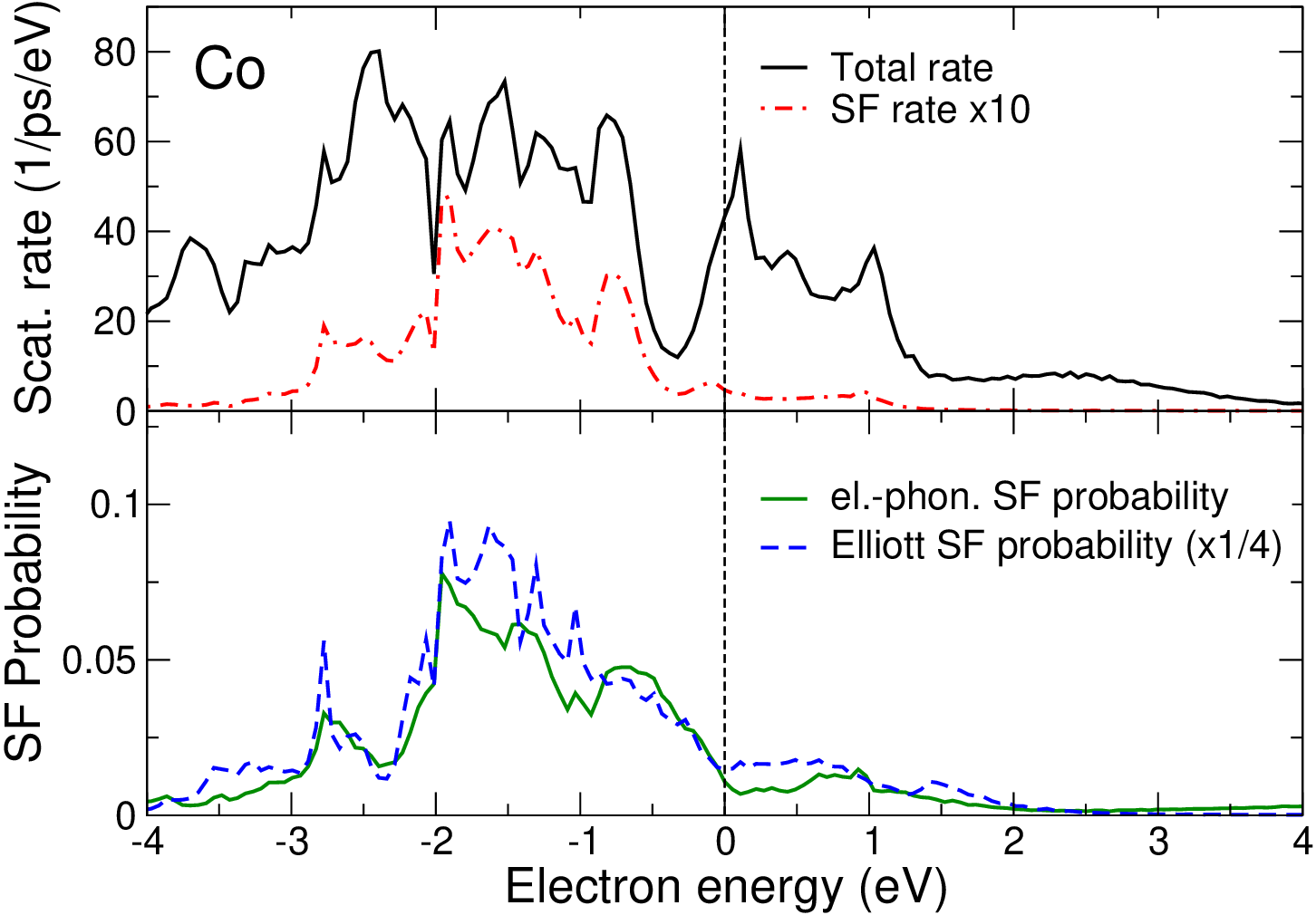}
\caption{(color online) \textit{Ab initio} calculated energy-resolved  total and
SF electron-phonon scattering rates of fcc Co (top), and
calculated energy-resolved SF probability
as well as the SF probability (divided by four)
obtained from Elliott's relation (bottom).}
\label{fig:En-rates-Co}
\end{figure}

The calculated energy-resolved electron-phonon scattering rates and SF probabilities of Co
 are given in Fig.\ \ref{fig:En-rates-Co}.
The SF scattering rate around $E_F$ is here particularly smaller than in Fe and Ni, whereas
the shape of the energy-dependent SF rate 
%(Fig. \ref{fig:En-rates-Co})
is intermediate to those of Ni and Fe, with a smaller peak above the Fermi level.
We further note that the total electron-phonon scattering rate exhibits a deep minimum below $E_{F}$ and maximum
above it---features that do not simply correspond to the total DOS. 
The directly computed energy-resolved SF probability (Fig. \ref{fig:En-rates-Co}, bottom)
is lower than in Ni and Fe; this gives a SF rate which is a 100 times smaller than the total rate near $E_F$. For fcc Co we furthermore
find that the structures in the directly computed
SF probability and that obtained from Elliott's relation agree relatively well, however, Elliott's SF probability systematically
overestimates the true SF probability by a factor four.
This indicates that, depending on the studied material, the agreement between the two energy-resolved SF probabilities 
can range from being reasonable to poor, with Elliott's SF probability being typically larger by factors of two to four.

%\begin{figure}
%\includegraphics[width=0.85\columnwidth]{fig_Co/Co_SFRates}
%\caption{\label{fig:En-rates-Co} (color online) Energy-resolved electron-phonon total and
%spin-flip (SF) scattering rates $w^{ep}$ and $w_{S}^{ep}$, SF probability
%$p_{S}^{ep}$ and (approximate SF probability from Elliott relation
%$p_{SF}^{b^{2}}$ ) for Co.}
%\end{figure}

\subsection{Nonequilibrium hot electron spin lifetimes }

The lifetimes as well as the spin lifetimes of excited nonthermal electrons due to electron-phonon
scattering can be computed within our approach.
To obtain these lifetimes at a given electron energy $E$ above the  Fermi energy, we define them
 as the average over all states having energy $E$.
The average electron lifetimes correspond
to the inverse scattering rate per number of states, i.e.,
 $\tau_{el}^{\sigma} (E)=n_{\sigma}(E)/{ \sum_{\sigma\sigma'} w_{\sigma\sigma'}(E)}$,
while the spin lifetimes are given as 
\begin{equation}
\tau_{_{SF}}^{\sigma}(E)=n_{\sigma}(E)/w_{\uparrow\downarrow}(E)\,.
\end{equation}
 We assume that all other states above $E_{F}$ are available for the NEQ electron to scatter
into. 
%{\red a value of e-p tau? KC: a graph can be added}\\

\begin{figure}
\includegraphics[width=0.8\columnwidth]{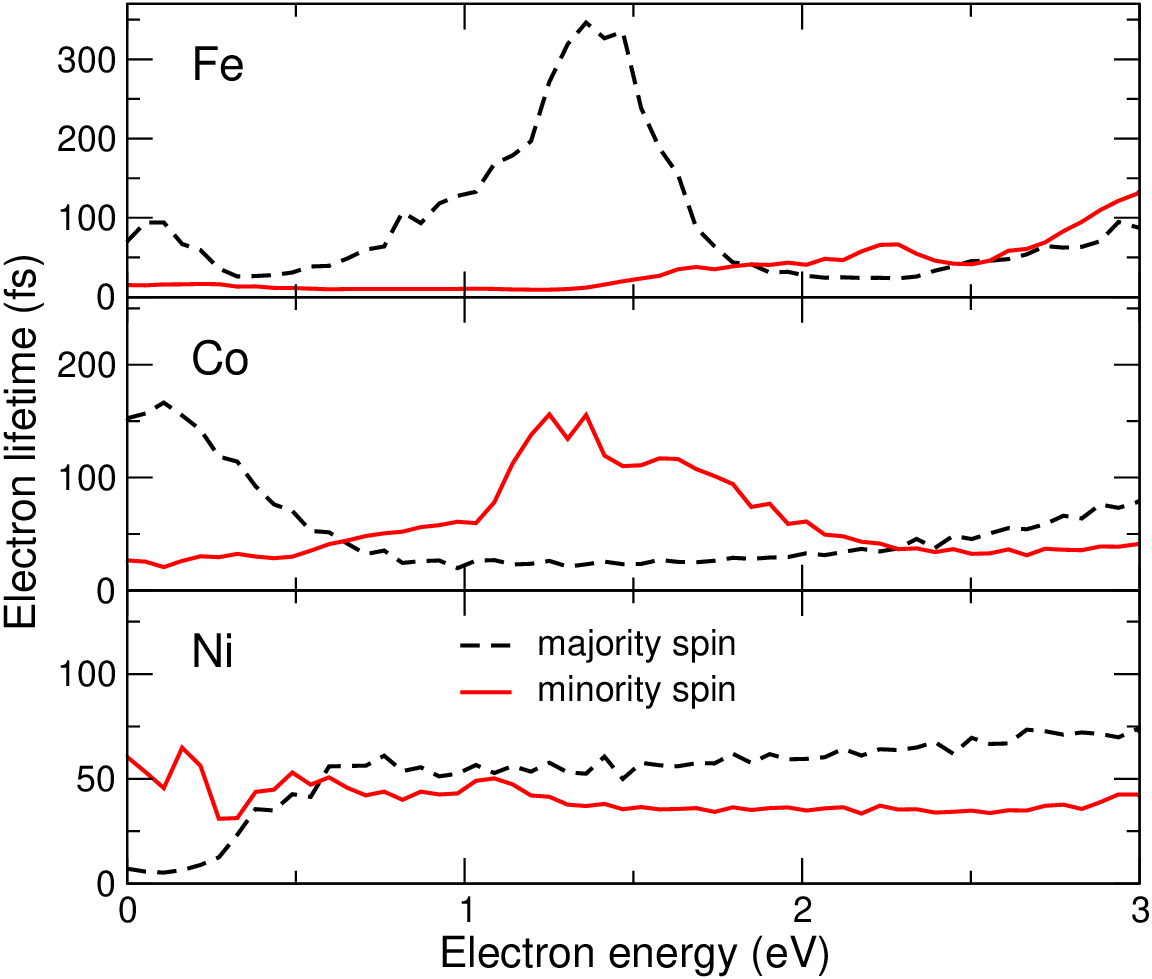}
\caption{\label{fig:El-lifetime} (color online)
\textit{Ab initio} computed spin-resolved electron lifetimes $\tau_{el}^{\sigma} (E)$ caused by electron-phonon scattering of hot electrons  in  Fe, Co, and Ni, given as a function of the hot electron energy relative to the Fermi level at 0 eV.}
\end{figure}

The calculated spin-resolved lifetimes $\tau_{el}^{\sigma} (E)$ of hot electrons due to (spin-conserving) electron-phonon scattering of Fe, Co, and Ni are
shown in Fig.\ \ref{fig:El-lifetime}.
%The here-computed total electron-phonon lifetimes are on a comparable timescale as those
%that are due to electron-electron scattering.\cite{r_06_Zhuko_IMFP_GW_FeNi}
%%{\red how is this meant? there is a factor 1000 ? -P KC:not spin lifetimes, electron lifetimes}
In some cases the calculated electron-phonon lifetimes are on a comparable timescale as those 
due to electron-electron scattering.\cite{r_06_Zhuko_IMFP_GW_FeNi} This is most evident for minority spin electrons in Fe for energies less than 1.5 eV, where we find hot electron lifetimes of around 10\,fs for electron-phonon scattering, while very similar values have been predicted for electron-electron scattering.\cite{r_06_Zhuko_IMFP_GW_FeNi} In this situation both processes contribute almost equally, thus halving the effective total lifetime, while for majority electrons the contribution coming from electron-phonon scattering is small. This observation could contribute to explain the observed spin asymmetry of  the hot-electron mean free path in Fe, \cite{Banerjee07} which does not follow the expectation based on \textit{ab initio} computed\cite{r_06_Zhuko_IMFP_GW_FeNi} electron-electron lifetimes.
We find that the electron-phonon lifetimes can be quite energy as well as spin dependent, in particular for Fe and Co. The electron-phonon induced  lifetime $\tau_{el}^{\sigma}$ of Fe has a huge \textit{spin asymmetry} for electrons with energies in the interval 0.7 to 1.7 eV, i.e., in the range of energies that can be reached by laser excitation. {This is related to the \textit{spin-resolved} scattering rates (not shown here).} 
 The strong energy dependence of the spin-resolved lifetimes occurs at energies where the $d$ character of the bands prevails. At higher energies, where the hot electrons have mainly $sp$ character the electron-phonon lifetimes of majority and minority spin electrons are relatively comparable, having spin-averaged values of about 50 fs in Co and Ni, and up to 100 fs in Fe.

\begin{figure}
\includegraphics[width=0.8\columnwidth]{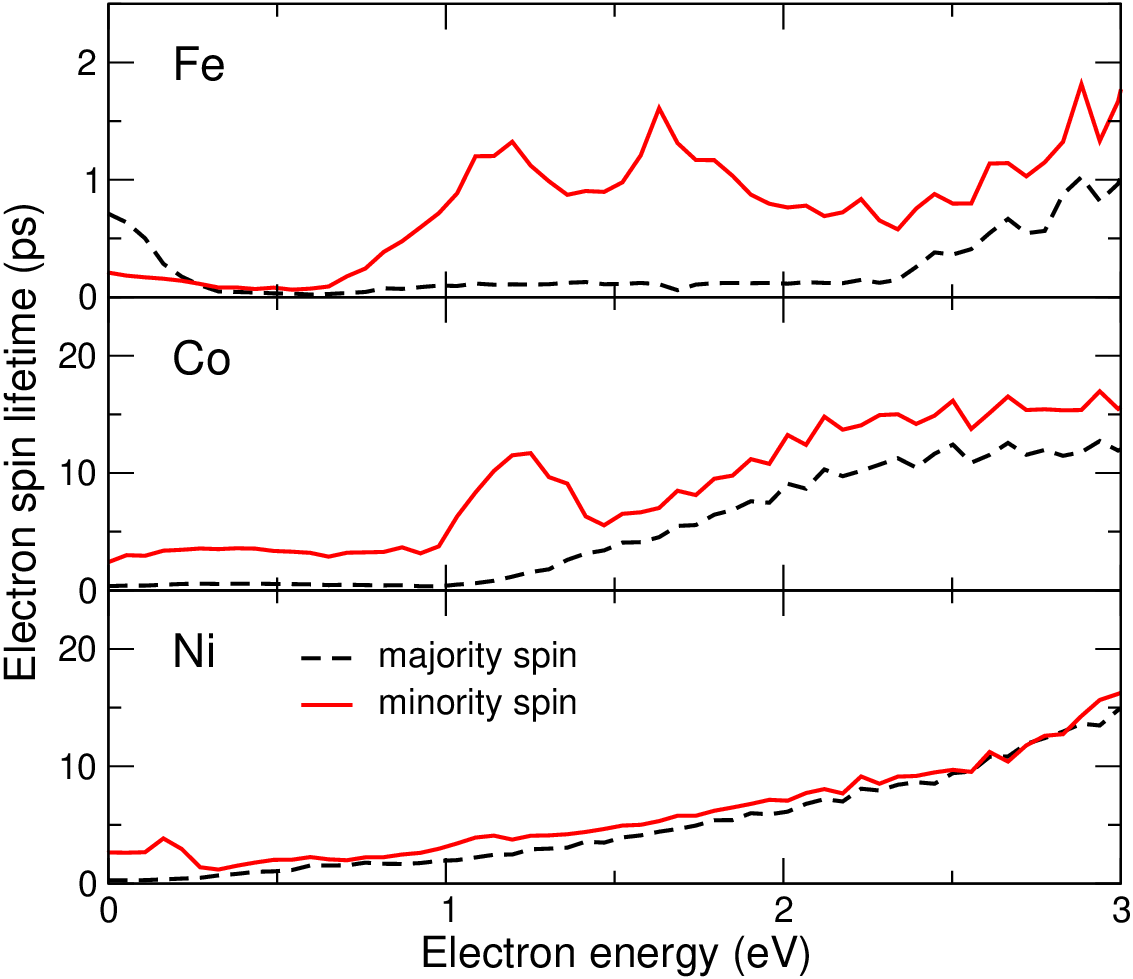}
\caption{\label{fig:El-spin-lifetime} (color online)
Computed  spin-flip lifetimes $\tau_{_{SF}}^{\sigma}(E)$ of spin-polarized hot electrons due to
electron-phonon scattering in Fe, Co, and Ni, given as a function of the electron energy relative
to the Fermi level at 0 eV.}
\end{figure}

In Fig.\ \ref{fig:El-spin-lifetime} we show the \textit{ab initio} computed electron-phonon induced electron spin lifetimes $\tau_{_{SF}}^{\sigma}(E)$
for all three $3d$ ferromagnets as a function of the nonthermal electron energy.
 A notable and common feature is the increase for higher energies, which
 is due to the $s$- and $p$- characters of these states  becoming increasingly prominent. Simultaneously
 the degree of spin-mixing in these $sp$-states is lower and the SF scattering
rate therefore decreases.
A further interesting feature is that for bcc Fe the order of the spin-majority and minority lifetimes reverses below 
about 0.5 eV. A crossing of the Fe spin-majority and spin-minority lifetimes due to electron-electron
scattering has previously been predicted below 1 eV.\cite{r_06_Zhuko_IMFP_GW_FeNi}
The origin of this crossing has already been explained by phase space considerations, i.e., the amount of spin-majority and minority
phase spaces available for scatterings.\cite{r_06_Zhuko_IMFP_GW_FeNi}
The hot electron spin lifetime in Fe is considerably smaller than that in Co and in Ni. This is due to the much larger scattering rate in Fe at $1 -2$ eV electron energy, see Fig.\ \ref{fig:En-rates-Fe}.
%{\green why is there a factor 10 in Fe ? -more KC: because  Fe has a bigger scattering rate above E$_F$ compared to other two - this can be seen in the graphs of scattering rates }
%{\red the SF to normal lifetime behaves unusual in Fe -dashed curves?!}
The overall shape of the spin lifetimes also reflects the character of the unoccupied states.
In Fe there are unoccupied spin-majority $d$ states up to about 0.5 eV, whereas there are spin-minority $d$ states to about 2.5 eV.
These states are responsible for the more pronounced structures below these energies. Above these energies there are 
dominantly $sp$ states giving relatively flat, slowly increasing spin lifetimes.
In Ni there are only spin-majority $sp$ states present above $E_F$ and the spin-minority $d$ band extends to 0.3 eV,
concomitant with the energy dependences of the spin lifetimes seen in Fig.\ \ref{fig:El-spin-lifetime}.
Co adopts a position intermediate between Fe and Ni.

\subsection{Electron-phonon generated demagnetization rates 
\label{sub:Pumped-system}}

\subsubsection{Choice of hot electron distribution functions}

We now turn to the demagnetizing influence of phonon-mediated  SF scattering, particularly considering laser-heated ferromagnets.
The determining quantities are the spin-resolved transition rates, Eq.\ (\ref{eq:trans_rates}), and the demagnetization ratio $D_S$,
Eq.\ (\ref{eq:D_S}).
The spin-resolved transition rates depend strongly on the spin-dependent electron distributions
in the system. If $f_{\uparrow} (E )=f_{\downarrow} (E)$ the spin-increasing and spin-decreasing
scattering rates $S^+$ and $S^-$ would be identical and consequently
there would be no change of the total moment regardless of the value of the total SF probability,
$P_{S}$.  It is the imbalance between the scattering rates $S^+$ and $S^-$ which is essential for changing the moment 
(compare the two expressions in Eq.\ (\ref{eq:rates+-})). The condition of unequal spin distributions 
is provided 
{in ferromagnets with exchange-split spin-polarized bands, because the pump laser excitations will be different for the available spin-majority and minority bands which leads to generation of unequal spin populations. 
The laser excitation in itself is, to a good approximation, spin conserving, even in a full Dirac treatment with spin-orbit interaction.\cite{r_95_Kraft_PMO_AccurDirac_MOKE}}
 The occurring spin (and charge) conservation can be imposed on the distributions by requiring 
\begin{equation}
\int dE f_{\sigma} (E) n_{\sigma} (E) = \int dE f_{\sigma}^{0} (E )n_{\sigma} (E)\,,
\label{eq:TotalSpinCon}
\end{equation}
where $f_{\sigma}^{0}$ is the low-temperature Fermi-Dirac distribution.

Immediately following irradiation with the pump laser the initial
electron population is in a nonthermal state far from any Fermi-Dirac
distribution. The  excited hot electrons in this state thermalize by electron-electron scatterings, 
a process which leads within approximately 300 fs to a thermalized electron distribution.
\cite{sun93, r_00_DelFa_NEQPhDynNobleM, guo01, rhie03, Lisowski04}
%\cite{r_00_Koopm_StateFill,r_00_DelFa_NEQPhDynNobleM,r_00_Hohlf}
The thermalized electron distributions are captured by Fermi-Dirac distributions with a high, spin-dependent electron temperature $T_{\sigma}$.
As mentioned earlier, these hot, thermalized electrons are not yet in equilibrium with the lattice, \cite{r_10_Wang_TempDepElPhon} therefore we assume
 the phonons to be those of the equilibrium system.
% fast  and within 100-300 fs it thermalizes by elelectron-electron scattering. 
We further note that any spin relaxation occurring during electron-electron scattering
is not treated here and has been studied elsewhere.\cite{r_08_Zhuko_GW_SpinRelaxExc_SO,r_09_Krauss_CoNiComp_Coulomb}

The nonthermal\cite{r_09_clo_NiDemag} and thermalized
electron distributions are computed here for similar total absorbed laser energies.
The Fermi-Dirac distribution in the thermalized situation is given by a $f_{\sigma}(E)=f_{FD}(E,\, T_{\sigma},\, \mu_{\sigma})$; on account
of the laser imparted energies electron temperatures of the order of a few thousands $\mathrm{K}$ can
be expected.
% as long as the system is not equilibrated with the lattice.
At such temperatures the spin-dependent chemical potentials $\mu_{\sigma}$ are shifted from their equilibrium value 
and have to be determined by solving  Eq.\ (\ref{eq:TotalSpinCon}) numerically.
 Note that the shift in $\mu_{\sigma}$ has to go beyond the Sommerfeld expansion because the density of states changes significantly
within the energy range set by the high electron temperature (3000 K corresponds to 0.26 eV). 
%Therefore the chemical potentials $\mu_{\sigma}$ are found numerically by solving eqs. (\ref{eq:TotalSpinCon}).

The electron distributions present before thermalization are more difficult
to describe. In principle a non-monotonic band, spin, and time-dependent
occupation function $f_{b,\sigma}(E,t)$ would be needed, but
such a precise description has not yet been achieved for real materials.
Here we therefore concentrate on its main features, which can be obtained from
an optical conductivity calculation to simulate the influence of the pump laser. 
The influence of the pump laser on the total number of valence electrons is 
still relatively small, therefore the induced distribution can
be given as an equilibrium distribution plus a deviation,\cite{r_00_Hohlf,r_04_Oppen_NiDemag}
  when assuming the usual fluences, which also implies that we do not have to consider any change
in the band structure.
The energy boundaries of the deviation are set
by the frequency of the laser, inside this energy window the distribution
is assumed to be flat. The critical component for magnetization dynamics
is the spin-dependence of this distribution, see Eq.\ (\ref{eq:rates+-}), therefore we concentrate
on this feature. 

The laser-created  NEQ distributions are determined as follows.
From $f_{exc}$ excited electrons $\gamma_{\uparrow}f_{exc}$
are spin-up and $\gamma_{\downarrow}f_{exc}$ are spin-down electrons, where
$\gamma_{\uparrow}+\gamma_{\downarrow}=1$. The ratio between excited
spin-majority and minority electrons $\gamma_{\uparrow}/\gamma_{\downarrow}$
is obtained from the \textit{ab initio} calculated spin-resolved optical conductivity $\mathrm{Re}\,[\sigma_{xx}(\omega)]$,\cite{r_04_Oppen_NiDemag, Hild12}
where the exciting laser frequency $\omega=1.55$ eV is assumed. The
quantities related to the regions $[E_{{F}}-\hbar\omega,\, E_{{F}}]$
and $[E_{{F}},\,E_{{F}}+\hbar\omega ]$ are marked with superscripts
$<$ and $>$, respectively. Employing the total numbers of spin-majority and
minority electrons in these regions, $N_{\sigma}^{<}$ and $N_{\sigma}^{>}$, 
one obtains the spin-dependent occupation factors relative to $f_{exc}$: $f_{\sigma}^{<}=(\gamma_{\sigma}/N_{\sigma}^{<}) f_{exc}$
and equivalently for $f_{\sigma}^{<}$.
The \textit{ab initio} computed numbers  for Ni, Fe, and Co 
are given in Table \ref{tab:SpinOccup}.
%{\red should it be 1.5 or 1.59 eV ? -P}

\begin{table}[t,b]
\begin{ruledtabular}
\caption{\label{tab:SpinOccup} Given are the 
spin-decomposition of the optical transitions $\gamma_{\sigma}$
induced by an $\omega$=$1.55$ eV laser, number of electrons $N_{\sigma}^{<}$ and $N_{\sigma}^>$
accommodated in the energy windows $[E_{{F}}-\hbar\omega,\,E_{{F}}]$ and
$[E_{{F}},\,E_{{F}}+\hbar\omega]$  as well as their
resulting occupations $f_{\sigma}^{<}$ and $f_{\sigma}^>$.}
\begin{tabular}{>{\centering}p{0.15\columnwidth}>{\centering}p{0.15\columnwidth}>{\centering}p{0.15\columnwidth}>{\centering}p{0.15\columnwidth}>{\centering}p{0.15\columnwidth}>{\centering}p{0.15\columnwidth}}
%\hline 
\noalign{\vskip\doublerulesep}
atom,$\,\sigma$ & $\gamma_{\sigma}$ & $N_{\sigma}^{<}$ & $f_{\sigma}^{<}/f_{exc}$ & $N_{\sigma}^{>}$ & $f_{\sigma}^{>}/f_{exc}$\tabularnewline[\doublerulesep]
\hline 
\noalign{\vskip\doublerulesep}
Ni$\,\uparrow$ & $36$\% & $1.65$ & $0.22$ & $0.22$ & $1.65$\tabularnewline
Ni$\,\downarrow$ & $64$\% & $1.38$ & $0.46$ & $0.69$ & $0.92$ \tabularnewline[\doublerulesep]
Fe$\,\uparrow$ & $29$\% & $2.11$ & $0.14$ & $0.35$ & $0.83$\tabularnewline
Fe$\,\downarrow$ & $71$\% & $1.43$ & $0.50$ & $1.34$ & $0.53$\tabularnewline[\doublerulesep]
Co$\,\uparrow$ & $28$\% & $1.62$ & $0.17$ & $0.24$ & $1.19$\tabularnewline
Co$\,\downarrow$ & $72$\% & $1.66$ & $0.44$ & $1.67$ & $0.43$\tabularnewline
%\hline 
\end{tabular}
\end{ruledtabular}
\end{table}

%{\red KC: I changed this and provided more robust argument because it is not completely valid for Fe with its minimum of minority DOS near $E_F$, and it should be merged with discussion of results further down: }\\
%{\cyan do you mean that the previous text should completely go out? Also, you talk about demagnetisation, but this is first discussed in the next paragraph - so I put it there - P}

%To understand the significantly weaker demagnetization in the thermalized regime one may note that in all studied metals the highest SF rates were found to be positioned at least 0.5 eV below the Fermi level. However, in the electron thermalized situation for the here-studied temperatures the difference between the Fermi-Dirac distributions at high-temperature and at zero temperature almost vanishes for energies below 0.5 eV (reaches more than 0.9 for 3000~K) 
%{\red is not 5000K almost 0.6 eV? -P}. 
This significantly limits SF contributions from the high binding energy region, where a high number of spin-flips take place in the case of non-thermal electron distributions. Furthermore, the largest imbalance between minority and majority occupation numbers  appears at energies above $E_F$. \cite{r_11_cbo_EYSF} 

%{\blue Perhaps:\\
%This however implies that for electrons
%below the Fermi level spin-decreasing flips will be fewer, therefore an overall demagnetization
%can occur only through the counteracting influence of electrons above the chemical potential. KC: better replacement above. }

\subsubsection{Computed results}

The main results of the electron-phonon induced SF scattering are collected in Table \ref{tab:SFRates-1}.
To start with, the Elliott SF probabilities $P_S^{b^2}$ computed in this work are compared with previously computed  Elliott
probabilities of Steiauf and 
F{\"a}hnle,\cite{r_09_Steia_Fahnl_EYcalcul} as well as with the exact electron-phonon SF probabilities 
$P_S$. Our calculated Elliott probabilities are comparable ($\sim 30\%$) with the earlier computed ones; a somewhat larger deviation
of 36\% is present for fcc Co.\cite{r_09_Steia_Fahnl_EYcalcul} 
The deviation between the Elliott SF probability $P_S^{b^2}$ and the true SF probability is notably larger, ranging from 
a factor of two to six. The deviation is particularly large for fcc Co. 
Contrary to previous results (based on the Elliott approximation) \cite{r_09_Steia_Fahnl_EYcalcul} for Co,
we predict a SF probability $P_S$  of Co that is four times lower than that of Ni, also in
the excited regime.

Steiauf and F{\"a}hnle\cite{r_09_Steia_Fahnl_EYcalcul} have estimated the Elliott probability $P_S^{b^2}$ (denoted as $\alpha_{\textit{sf}}$ in Ref.\ \onlinecite{r_10_Koopm_EYDemagInclGd}) for the laser-excited ferromagnets Ni and Co by averaging the energy-dependent spin-mixing probability $ \langle b^2 (E) \rangle$ using a Gaussian with standard deviation $\sigma = 1.4$ eV around $E_F$. They obtained thereby enhanced probabilities $\alpha_{\textit{sf}} = P_S^{b^2}$ of 0.18 and 0.196 for Ni and Co, respectively. These values cannot be directly compared to our computed values for the laser-irradiated materials in Table \ref{tab:SFRates-1} as there we have used explicitly NEQ and thermalized electron distributions to simulate the effect of the pump laser. It may be noted, however, that the difference with the exact SF probability $P_S$ can be large, up to a factor of 10 for Co.

Considering next the electron-phonon SF probabilities $P_S$ in comparison with the demagnetization ratios $D_S$,
we observe that the latter are substantially smaller than the former. For a laser-heated thermalized electron gas the demagnetization ratios $D_S$ are
about ten times smaller than $P_S$, exemplifying that there can be a moderately high total SF probability, which, however with nearly as many spin-increasing as spin-decreasing  SF events does not give rise to an appreciable demagnetization. 

\begin{table}
\begin{ruledtabular}
\caption{\label{tab:SFRates-1}
Given are \textit{ab initio} calculated spin-flip probabilities $P_S$,  Elliott SF probability $P_{S}^{b^{2}}$, demagnetization ratios $D_S$, and relative demagnetization fractions $\Delta M /M_0$ for laser-pumped Ni, Fe, and Co. Calculated values are given for equilibrium (low $T$), 
for thermalized electrons at a high Fermi temperature $T_e$, and for the nonequilibrium (NEQ) electron distribution created by fs laser-excitation. Computed values for the approximate Elliott SF probability $P_{S}^{b^{2}}$ are
compared to values from Ref.\ \onlinecite{r_09_Steia_Fahnl_EYcalcul}.
The relative demagnetization fraction $\Delta M /M_0$, achieved by electron-phonon SF scattering, is given in $\%$ at 250 fs.}
\begin{tabular}{l c c c c}
%\hline 
 & $P_{S}^{b^{2}}$ & $P_{S}$ & $D_{S}$ & $\Delta M/M_0$\tabularnewline[\doublerulesep]
\hline 
\noalign{\vskip\doublerulesep}
Ni (low $T$) & 0.07 (0.10 \cite{r_09_Steia_Fahnl_EYcalcul}) & $0.04$ & 0 & 0\tabularnewline
%\hline 
Ni ($T_{e}$=3000\,K) & 0.11 & 0.07 & 0.003 & 3.1\tabularnewline 
%\hline 
%Ni ($T_{e}$=5000\,K) & 0.12 & 0.1 & 0.004 & \tabularnewline
%\hline 
Ni (NEQ) & 0.12 & 0.09 & 0.025 & 
%(0.8) 
16.7 \\[0.2cm]
%\tabularnewline[\doublerulesep]
%\hline 
Fe (low $T$) & 0.068 (0.096 \cite{r_09_Steia_Fahnl_EYcalcul})  & 0.04 & 0 & 0\tabularnewline
%\hline 
Fe ($T_{e}$=3000\,K) & 0.13 & 0.09 & 0.008 & 
%(0.4)
4.5\tabularnewline
%\hline 
Fe (NEQ) & 0.14 & 0.07 & 0.030 & 
%(1.9)
11.4 
%{\red 22} 
\\[0.2cm]
%\tabularnewline[\doublerulesep]
%\hline 
Co (low $T$) & 0.060 (0.044 \cite{r_09_Steia_Fahnl_EYcalcul}) & 0.010 & 0 & 0\tabularnewline
%\hline 
Co ($T_{e}$=3000\,K) & 0.095 & 0.017 & 0.002 & 
%(0.06)
0.9\tabularnewline
%\hline 
Co (NEQ) & 0.105 & 0.022 & 0.010 & 
%(0.3)
2.3 \tabularnewline
%\hline 
\end{tabular}
\end{ruledtabular}
\end{table}

As noted previously \cite{r_11_cbo_EYSF} for fcc Ni, a somewhat larger SF probability but a drastically larger demagnetization ratio is 
obtained for the highly  NEQ state immediately after laser excitation. This is observed for all three ferromagnets.
The mechanism of the larger NEQ demagnetization ratio is complex as it involves hole and electron contributions, above and below $E_F$. 
To understand the significantly weaker demagnetization ratio in the thermalized regime one may note that in all studied metals the highest SF rates were found to be positioned at least 0.5 eV below the Fermi level. However, in the electron thermalized situation for the here-studied temperatures the difference between the Fermi-Dirac distributions at high-temperature and at zero temperature becomes small for energies  of more than 0.5 eV below $E_F$ (difference becomes less than 0.1 for 3000~K).
This significantly limits SF contributions from the higher binding-energy region, where a high number of spin-flips takes place in the case of non-thermal electron distributions. In addition, the largest imbalance between minority and majority occupation numbers  appears,
for all three ferromagnets in the electron thermalized situation, at energies above $E_F$. \cite{r_11_cbo_EYSF} 
Hence, we observe that SF electron-phonon scattering is efficient only as long as the
electron spin-distributions are in a highly NEQ state, with significant contributions coming from deep-lying hole states. 

The calculated electron-phonon mediated relative demagnetization fractions $\Delta M/M_0$, which are a central outcome of our investigation, are given in Table \ref{tab:SFRates-1}.  $\Delta M$ denotes $M_0 - M(\tau )$, and 
 $M_0$ is the computed static equilibrium moment; the values for $M_0$ are  2.2, 1.6, and 0.64 $\mu_B$ for Fe, Co, and Ni. 
 The resulting demagnetization fractions $\Delta M /M_0$ are expressed in percentage of demagnetization at time $\tau = 250$ fs. 
The time interval of 250 fs is chosen for the following reasons: laser-induced demagnetization experiments show that the process of demagnetization is completed at about 250 fs in ferromagnetic transition metals\cite{r_07_Stamm_FemtoNi, r_08_Carpe_ElMagnon_Fe, r_05_Hilto_ultraTHz_Fe, Stamm10} and their alloys.\cite{r_12_Mathi_ExchIntTime_NiFe} 
The computed relative demagnetizations can thus directly be compared with experimental data. Furthermore, photoemission experiments on the evolution of the laser-induced NEQ state show that this state evolves into an electron thermalized state within about the same time.\cite{sun93, guo01, rhie03, Lisowski04} 
Once the NEQ state has thermalized the large contribution to the SF ratio has vanished. As it is not precise known how the nonthermal state evolves to a thermalized one, we assume here for simplicity that the NEQ state decays linearly with time to a thermalized state in 250 fs. 
Conversely, for the calculations of $\Delta M /M_0$ in the electron thermalized state we assume that this state is stable over 250 fs, i.e, it is assumed that the electron thermalized state is formed instantaneously upon laser-excitation and then persists. This is an approximation, as there will be a transfer of energy from the hot electrons to the lattice. However, recent measurements showed that it takes about 3 ps for the lattice to heat up;\cite{r_10_Wang_TempDepElPhon} hence, its influence is estimated to be limited within 250 fs.

The calculated relative demagnetization fractions due to electron-phonon SF scattering are negligible at low temperatures, as could be expected. A more important outcome of our calculations is that at relatively high electron temperatures of 3000 K the achieved amount of demagnetization is only a few percent. In particular, the rather small computed values show that phonon-mediated SF scattering in the thermalized situation cannot explain the experimentally observed demagnetizations of the order of 50{\%}.\cite{r_96_Beaur_UltraNiDynam, Cheskis05, r_07_Stamm_FemtoNi, r_10_Koopm_EYDemagInclGd, Weber11}
 Alternatively, using $dM(t)/dt \large|_{t=0} = -[M_0-M_{\rm min.}]/\tau_M$ we can estimate the time $\tau_M$ needed to achieve a 50{\%} demagnetization,  $M_{\rm min.} = 0.5 M_0$. For an electron temperature of 3000 K this gives 3 ps, 14 ps, and 4 ps, respectively, for Fe, Co, and Ni. These demagnetization times computed for the electron-thermalized regime are considerably longer than the experimental ones.

The situation is quite different for NEQ electron distributions. These can give an appreciable contribution to the total demagnetization; notably, the resulting values are strongly material dependent. A moderately high relative demagnetization of 17{\%} is computed for Ni, whereas for Co it is only 2{\%}. These surprising  materials' dependences  are directly born out of the energy-resolved electron-phonon SF scattering rates which markedly depend on the spin-polarized electronic structures near $E_F$.  Irrespective, the total computed electron-phonon mediated demagnetizations do not reach the experimental demagnetizations of about 50{\%}.

%For Ni the SF scattering rate is very low in the region $E>E_{F}$,
%as shown in Fig.\ \ref{fig:En_SF}. 
%%This situation is demonstrated in Fig. 3 of \cite{r_11_cbo_EYSF}. 
%Due to these reasons the calculated
%demagnetization exhibits only a weak temperature dependence and is
%slow compared to the case below (Tab. \ref{tab:SFRates-1}).

The surprising materials dependencies can be explained in terms energy-dependent SF scattering rates and the particular shapes
of the spin-polarized DOS in each of the ferromagnets, but an extensive analysis is outside of the focus of the present study.
We briefly mention that, due to the $d$-band positions, the SF scattering rate in the region $E > E_F$ is very low for Ni, but much larger for Fe.
The demagnetization in the thermalized regime is therefore relatively small and only weakly temperature dependent in Ni, but in Fe 
it is not suppressed as strongly. In Fe the peak in the SF scattering rate above $E_F$ (see Fig.\ \ref{fig:En-rates-Fe}) contributes to the effective demagnetization rate through spin-decreasing transitions in the electron channel. This leads to the highest computed demagnetization gradient 
$dM /dt (t=0)$ of the studied materials, being counterbalanced by the highest static moment $M_0$.

%There are important features of Fe DOS (Fig. \ref{fig:Fe-DOS}) affecting
%spin-flips: Fermi level lies between two large peaks of the minority
%DOS and on an edge of a large decrease for the majority DOS. As a
%result again there is no strong imbalance favoring spin decreasing
%transitions in the region of the large peak of SF transition rate
%$w_{S}\left(E\right)$ centered at -1 eV below the Fermi level (Fig.
%\ref{fig:En-rates-Fe}), but for Fe there is another peak above the
%Fermi level, which, though weaker, allows for spin flips through electron
%channel. As a result the demagnetization rate in the thermalized case
%is not suppresed as strongly as in the Ni case.

%For NTED of Fe described in Tab. \ref{tab:SpinOccup} the SF probability
%is on average slightly lower than in Ni (see Tab. \ref{tab:SFRates-1}).
%Due to the strong average electron-phonon coupling and a higher imbalance
%in favour of spin descreasing transitions the demagnetization rate
%is highest of all the studied metals: $2\mu_{B}/$ps and atom. However
%Fe atoms have a high magnetic moment of $\mu_{Fe}=2.2$ $\mu_{B}$,
%hence the relative change rate is only $0.9\mu_{Fe}/$ps. \emph{Any
%comparable experimental data on Fe?}

\section{Discussion}

A paramount outcome of our {\it ab initio} investigation is that electron-phonon SF scattering processes in the electron thermalized regime cannot account for observed laser-induced demagnetizations. Before this can be definitely established it is however required, first, to critically discuss the approximations in our approach, and second, compare with other recent calculations.

Recently, Essert and Schneider \cite{r_11_Essert_EPScat_Demag} performed an \textit{ab initio} investigation, in which they carried out a direct electron-phonon matrix calculation, employing the rigid-ion approximation, to compute SF probabilities in fcc Ni and bcc Fe. 
They furthermore calculated the demagnetization rate employing the Boltzmann equation.
The computed thermalized demagnetization fractions are of the order of a few percent, in good agreement with our values.
This is furthermore consistent with our findings which show that Eliashberg functions computed with the rigid-ion approximation agree quite well with  Eliashberg functions computed without such approximation (Fig.\ \ref{fig:Elbg_SFElbg_Ni-1}).

Estimations of the effectiveness of the electron-phonon Elliott-Yafet SF contribution in the laser-induced demagnetization were made recently,\cite{r_10_Koopm_EYDemagInclGd, r_12_Roth_TempDepDemagNi_Mechanism} employing the microscopic three temperature model (M3TM) \cite{r_08_Longa_Thesis, r_10_Koopm_EYDemagInclGd} as well as computed Elliott SF probabilities, $P_S^{b^2}$. 
The main conclusion drawn in these investigations was that electron-phonon SF scattering in the electron thermalized regime can very well account for the measured demagnetizations and that the influence of highly excited electronic states can be disregarded.
Remarkably, our {\it ab initio} investigations lead to precisely opposite conclusions---the effective demagnetization reached in a thermalized electron system is quite small ($\sim 3\%$), but appreciably larger values result from NEQ excited distributions.
%{\red \sout{Notably, another recent {\it ab initio} investigation, \cite{r_11_Essert_EPScat_Demag} employing the rigid-ion approximation, computed thermalized demagnetization rates of a few percent, in good agreement with our values.}}
Hence, an important question to be addressed is why the M3TM predicts much larger phonon-mediated demagnetization values than {\it ab initio} based calculations, which will be done further below.
%It thus appears that {\it ab initio} calculations of Elliott-Yafet scattering consistently predict rather small demagnetization values, whereas the M3TM in contrast predicts much larger values.

Discussing now the approximations made in our approach, we start with noting that it is 
 essential to distinguish between transversal and longitudinal
magnetization relaxation. Transversal spin excitations are important
for the relatively slow magnetization precession on the pico- to nanosecond
scale, whereas longitudinal spin excitations might provide a channel
for sub-picosecond demagnetization. We have excluded the transversal
spin excitations in our calculations since they are not a part of
the common definition of Elliott-Yafet SF scattering.\cite{r_63_Yafet_SpinLatRel}

First, {\it a priori} it is not evident if our starting point, the Kohn-Sham LSDA band structure, nominally valid at $T=0$, is sufficient
to describe spin excitations in the $3d$ ferromagnets. It has been argued\cite{Schellekens13} that such a ``rigid" band structure cannot be 
used to describe demagnetization due to spin excitations.
%We have employed the Kohn-Sham band structure in our calculations.
It may be noted that it has been shown \cite{r_Halil_97, r_98_Halil_pmo_SpinWaveDFT} previously that
such zero-temperature band structure can very well be employed to accurately compute
the transversal spin-excitation energy spectra of Fe, Co, and Ni, as well as
their temperature-dependent magnetization, $M(T)$. However, as prior to our
calculations no \textit{ab initio} calculations of the longitudinal
Elliott-Yafet electron-phonon spin-flip spectrum existed, we can presently not state how accurate
such calculations are, but we may note 1) that our calculation for Al is in good agreement with an earlier work \cite{r_99_Fabia_Sarma_PhIndSpinRel} and 2) the recently published independent computational investigation \cite{r_11_Essert_EPScat_Demag} obtained very similar demagnetization fractions in the 
thermalized regime. 

Second, our calculations pertain to demagnetization effects in the linear
regime, i.e., at times immediately after the action of the laser pulse when the demagnetization
sets in. We do not study a complete time evolution, but we obtain
the magnetization's time derivative at $t=0$, $dM(t)/dt \large|_{t=0} = -[M(0)-M_{\rm min.}]/\tau_M$,
giving the demagnetization
rate.  %[i.e., {\red $dM(t)/dt \large|_{t=0} = -[M(0)-M_{\rm min.}]/\tau_M$]. 
This is consistent with the commonly-used 
 fit function $M(t) - M(0) \approx -[M(0)-M_{\rm min.}](1 - e^{-t/{\tau_M}})$ 
 %to obtain the experimental demagnetization time $\tau_M$ 
 (note that the re-magnetization happening on much longer times is neglected here).
A consequence of the here-assumed linear regime is that it does not take the effect of already occurred spin-flips
on the spin-polarized band structure into account. It could be that the demagnetization might accelerate 
cumulatively with the number of SF scattering events.\cite{r_12_Roth_TempDepDemagNi_Mechanism}
If this would be the case, the  time-dependent magnetization $M(t)$ would be a concave function.
However, $M(t)$ is always observed to
be a convex function in experiments and simulations. Thus,  extrapolating
$M(t)$ from its linear behavior at $t=0$ we would always be only overestimating
the effective demagnetization. 
 Furthermore, the peak achieved demagnetization has been observed experimentally to depend linearly on the fluence in a broad range of fluences, \cite{Cheskis05,r_12_Roth_TempDepDemagNi_Mechanism} which further justifies using the linear regime and the neglect of second order effects like that of already occurred spin flips.

 Third, it has been recently argued \cite{Essert12, r_12_Roth_TempDepDemagNi_Mechanism, Schellekens13}  
that in order to achieve a sizeable ultrafast demagnetization one would need to take into account an additional mechanism, besides the Elliott-Yafet electron-phonon SF scatterings,
for instance an ultrafast dynamical reduction of the exchange splitting. As mentioned above, through already occurred SF events the exchange splitting  would be somewhat reduced, however, this reduction is quite small initially and it happens on time scales that are relatively long compared to that observed for femtosecond demagnetization. Moreover, as the origin of exchange splitting is the Pauli exclusion principle it is not clear how an ultrafast  dynamic reduction can be achieved.

Fourth, we 
%have to 
note that the spin-flip processes studied in our model require significantly lower energy than Stoner excitations, because the SF scattering changes the electron energy only negligibly. The spin-flipped electron enters an empty state (hole) left by another previously laser-excited electron, or alternatively the spin-flipped hot electron had already acquired a high energy before undergoing a spin-flip. The necessary energy has thus already been gained from the pump laser and in contrast to Stoner excitations it is not needed for enabling the spin flip.

Next, in order to understand why our conclusions regarding the efficiency of the Elliott-Yafet SF mechanism are so
 different from recent evaluations on the basis of the M3TM model\cite{r_10_Koopm_EYDemagInclGd} 
we analyze more thoroughly the assumptions underlying the M3TM. 
In the M3TM  there is a disputable separation of electronic and spin degrees of freedom adopted; \cite{r_08_Longa_Thesis} the equilibration of the spin system with the lattice is driving the demagnetization. Further, the sharply peaked structures of the spin-polarized DOS, typical for each of the transition metals, have been assumed to be constant.
% Here the approach of
%Koopmans \cite{r_10_Koopm_EYDemagInclGd} employs several assumptions,
%first of all it is a disputable separation of spin and electronic
%degrees of freedom \cite{r_08_Longa_Thesis} and it does not take into account the peculiar DOS of Ni near the Fermi level. 
%
The difference between our computed exact SF probabilities $P_S$ and the Elliott SF probabilities $P_S^{b^2}$ used
\cite{r_10_Koopm_EYDemagInclGd} 
in the M3TM is a factor of 2 for Ni which may not be as significant as to lead to opposite conclusions, but it is a factor 10 for Co.
The crucial quantity which we identified for an ultrafast demagnetization\cite{r_11_cbo_EYSF}
is the imbalance between spin-increasing and spin-decreasing spin-flips which
correspond directly to the demagnetization rate. 
The essential difference between the M3TM and our calculations appears when this quantity
is computed, in the second step of the calculation to obtain $dM/dt$.
Taking a closer look at how this imbalance is provided in the M3TM, we note that the spin change rate $d{S}/dt$ is expressed as SF probability $a_{\textit{sf}}$ multiplied with transition factors which contain, e.g., $[1-f(E)]f(E + \Delta_{ex} + E_{ph})$, where $f$ is the Fermi function at an electron temperature $T_e$, $\Delta_{ex}$ the exchange splitting, and $E$ and $E_{ph}$ are the electron and phonon energies, respectively; see Eqs.\ (5.10) and (5.20) in 
Ref.\ \onlinecite{r_08_Longa_Thesis}.
The assumption behind this is that when a spin is flipped 
in the M3TM (in which electron and spin systems are separated)  the spin system gains energy $\Delta_{ex}$ and electron system looses
energy $\Delta_{ex}$, see Fig.\ 5.2 of Ref.\ \onlinecite{r_08_Longa_Thesis}. 
The spin-flipped electron is thus assumed to have
 a changed energy, e.g., $E + \Delta_{ex} + E_{ph}$, which is inserted in the Fermi-Dirac distribution.
In our calculations we used instead transition factors $[1-f_{\sigma}(E)]$ $f_{\sigma'} (E + \hbar \Omega)$, see Eq.\ (\ref{eq:SpinEvGen}), which express that when an electron scatters from one state to another, the two states can have energies different up to the phonon energy $\hbar \Omega$ (=$E_{ph}$). The M3TM conversely assumes that the electron scatters with a phonon to a spin-reversed state having a very large energy difference $\pm \Delta_{ex} \pm E_{ph}$ together with an energy gain $\mp \Delta_{ex}$ of the spin system. 
% This is unphysical, 
{We note that, as the exchange splitting $\Delta_{ex}$ ($\approx 1-2$ eV) is much larger than the typical phonon energy ($\sim 30$\,meV), it is not very probable that an electron can change its energy by $\Delta_{ex}$ in the electron-phonon scattering process.}
Employing such a massively changed electron energy in products of the form $[1-f(E)]f(E + \Delta_{ex} + E_{ph})$ will lead to spin-dependent transition rates that are vastly bigger than those 
obtained with $[1-f(E)]f(E + \hbar \Omega) \approx [1-f(E)]f(E)$,  as done in our approach. 
Consequently, because of the assumption made in the M3TM model the transition factors {are overestimated. 
This explains the difference of results between our model and M3TM (up to a factor of 50). Nonetheless, drawing a final conclusion about the demagnetization mechanism is still not possible at this stage, because there are several effects not present neither in the M3TM not in our model (for example, possible modification of band structure, exchange splitting and exchange constants due to the pump pulse), and in addition the electron redistribution following the pulse is  not accurately described in both models.} 
% unrealistically large, and accordingly, the estimated electron-phonon demagnetization \cite{r_10_Koopm_EYDemagInclGd} becomes substantially overestimated ($\sim$50 times larger than our result).

\section{Conclusions}

We have developed a theoretical treatment on the basis of a generalized spin- and energy- dependent Eliashberg function
to investigate computationally the Elliott-Yafet electron-phonon mediated ultrafast spin relaxation in ferromagnetic transition metals.
The formalism has been used---in combination with full-potential relativistic electronic structure
calculations as well as \textit{ab initio} computed phonon dispersions---to obtain materials' specific, quantitative
information with respect to the extent of ultrafast demagnetization that can be caused by spin-flip processes generated by electron-phonon scatterings in the $3d$ transition metals Fe, Co, and Ni. Our investigations specifically aim at clarifying the possible contribution of
Elliott-Yafet phonon-mediated spin-flip scatterings to the experimentally observed laser-induced ultrafast demagnetization. As yet,
these processes are still poorly understood, especially on the \textit{ab initio} level.
Our calculations reveal interesting and unexpected differences in the spin-relaxation behaviors of the three materials.
The calculated electron-phonon SF probability is observed to be strongly dependent on the electron single-particle energy. Also, the computed
energy- and spin- dependent hot electron
lifetimes and spin lifetimes exhibit considerable, unanticipated, differences for the three $3d$ ferromagnets. The electron lifetime of a hot electron in Fe has, for example, a much larger spin asymmetry than a hot electron in Ni.

As it has been proposed\cite{r_10_Koopm_EYDemagInclGd} that electron-phonon spin-flip scatterings in the electron thermalized regime could well account for the observed femtosecond demagnetization, we have computed the electron-phonon mediated demagnetization rate both for laser-created thermalized electron distributions as well as nonequilibrium (i.e., nonthermal) electron distributions. 
Our calculations aim at acquiring a more detailed understanding of the possible contribution
of Elliott-Yafet processes to the ultrafast demagnetization. We have
found that generally a difference of SF probabilities in states above and below Fermi level are important. 
Taking this energy-dependence of the \textit{ab initio} SF probabilities into account, 
we find that Elliott-Yafet SF processes in the electron thermalized regime cannot account for
the experimentally observed ultrafast demagnetization. Interestingly, the 
Elliott-Yafet phonon-mediatied mechanism is computed to be much more efficient (by a factor of 3 to 5) for nonthermal
laser-created electron distributions, as these enhance the imbalance of spin increasing and spin decreasing SF scatterings. 
Hence, we can conclude that the dominant contribution of the
Elliott-Yafet mechanism to ultrafast demagnetization has a nonthermal character and is expected to occur in the first few hundred femtoseconds, immediately after laser excitation.
The largest nonthermal Elliott-Yafet phonon-mediated demagnetization is computed for fcc Ni where it can reach about 17{\%} at 250 fs. 
While this is not negligible, it is not sufficient to explain the measured\cite{r_96_Beaur_UltraNiDynam, Cheskis05, r_07_Stamm_FemtoNi, Weber11} laser-induced demagnetizations of 50{\%} or more. {Summarizing, our calculations suggest that Elliott-Yafet} electron-phonon SF scattering alone cannot account for the observed ultrafast laser-induced demagnetizations;
{a different mechanism may be dominant here}. Recent experiments\cite{Rudolf12, Vodungbo12, Pfau12} suggest fast, nonequilibrium transport of spin-polarized hot carriers as a plausible source.

Lastly, we have compared our results with those previously obtained with the M3TM model. 
A significant difference between our approach and the M3TM model lies in the way the 
demagnetization rate is assigned with a given SF probability. As outlined above, we believe that this 
is done in a more accurate way in our theory than in the M3TM model.

\acknowledgments
{We acknowledge Christian Schneider, and Sven Essert and Manfred F{\"a}hnle for useful discussions. This work has been supported by the Swedish Research Council (VR), the G.\ Gustafsson Foundation, the Czech Science Foundation (P204/11/P481), and the European Community's Seventh Framework Programme (FP7/2007-2013) under grant agreements No.\ 214810 ``FANTOMAS" and No.\ 281043 ``FemtoSpin". Support through the Swedish National Infrastructure for Computing (SNIC) is gratefully acknowledged. National Center CERIT Scientific Cloud (CERIT-SC) Czech Republic is acknowledged for providing the computer facilities to support part of the present calculations under Grant Reg.No. CZ.1.05/3.2.00/08.0144. D. Legut acknowledges a support within the framework of the Nanotechnology Centre - the 
basis for international cooperation project, Reg. No. CZ.1.07/2.3.00/20.0074 and the IT4Innovations Centre of Excellence project, Reg. No. CZ.1.05/1.1.00/02.0070, both supported by Operational Programme 'Education for competitiveness' funded by Structural Funds of the European Union and state budget of the Czech Republic.}

\bibliographystyle{apsrev4-1}
\bibliography{b_kc-pmo}
%\bibliography{/Users/peter/EigeneDateien/Elliott/PRB/paper2/b_kc}

%{r_12_Mathi_ExchIntTime_NiFe}
%{r_08_Carpe_ElMagnon_Fe
%r_04_Tudos_UltSwitchLimit
%Bovensiepen09,

\end{document}